\documentclass[12pt,a4]{article}
\usepackage[T1]{fontenc}
\usepackage{lmodern}
\usepackage{bbm}

\usepackage{mathtools}
\usepackage{setspace} 
\usepackage[ansinew]{inputenc} 
\usepackage{amsthm}
\usepackage{mathtools}
\usepackage{commath}
\usepackage{float}
\usepackage{placeins}

\usepackage[T1]{fontenc} 
\usepackage{lmodern,microtype} 
\usepackage{graphicx}
\usepackage{titlesec,titling} 
\usepackage[nohead]{geometry} 
\usepackage{setspace} 
\usepackage{enumitem,booktabs} 
\usepackage{amsmath,amsfonts,amssymb,amsthm}
\usepackage{mathrsfs,ushort} 
\usepackage{graphicx,psfrag,epsf}
\usepackage{enumerate}
\usepackage{natbib}
\usepackage{bm}
\usepackage{dsfont}
\usepackage{url} 

\usepackage{hyperref}
\usepackage{bm}
\usepackage{mathtools}
\usepackage{dsfont}
\usepackage{comment}
\usepackage{breqn}
\usepackage{commath}
\usepackage{colonequals}
\usepackage{mathtools}
\usepackage{cases}
\usepackage{threeparttable}
\usepackage{mathtools}
\usepackage{graphicx}
\usepackage{empheq}
\usepackage{lmodern}
\usepackage{amsmath}

\usepackage{colonequals}
\RequirePackage{natbib}

\usepackage{threeparttable, tablefootnote}

\newcommand\Item[1][]{%
	\ifx\relax#1\relax  \item \else \item[#1] \fi
	\abovedisplayskip=0pt\abovedisplayshortskip=0pt~\vspace*{-\baselineskip}}

\DeclareSymbolFont{symbolsC}{U}{pxsyc}{m}{n}
\DeclareMathSymbol{\coloneqq}{\mathrel}{symbolsC}{"42}

\titleformat{\section}[block]{\centering\large\bfseries}{\thesection.}{0.5em}{}
\titleformat{\subsection}[block]{\flushleft\bfseries}{\thesubsection.}{0.5em}{}
\titleformat{\subsubsection}[runin]{\normalsize\itshape}{\bfseries\thesubsubsection.}{0.5em}{}[.--\:]

\titlespacing{\section}{0ex}{6ex}{3ex}

\usepackage[format=plain,justification=justified]{caption}

\begin{document}

\def\spacingset#1{\renewcommand{\baselinestretch}%
{#1}\small\normalsize}


  \title{\bf Serial-Dependence and Persistence Robust Inference in Predictive Regressions}
  \author{ Jean-Yves Pitarakis\thanks{I wish to thank the ESRC for its financial support via grant ES/W000989/1.  
  		Address for Correspondence: Jean-Yves Pitarakis, Department of Economics, University of Southampton, Southampton SO17 1BJ, United-Kingdom.}
  	 \\
    Department of Economics \\
    University of Southampton}
  \maketitle


\bigskip
\begin{abstract}
This paper introduces a new method for testing the statistical significance of estimated parameters in predictive regressions. The approach features a new family of test statistics that are robust to the degree of persistence of 
the predictors. 
Importantly, the method accounts for serial correlation and conditional heteroskedasticity without requiring any corrections or adjustments. This is achieved through a mechanism embedded within the test statistics that effectively decouples serial dependence present in the data. 
The limiting null distributions of these test statistics are shown to follow a chi-square distribution, and their asymptotic power under local alternatives is derived. A comprehensive set of simulation experiments illustrates their finite sample size and power properties.
\end{abstract}

\vspace{0.8cm}

\noindent%
\vfill

\newpage

\spacingset{1.32} 

\section{Introduction}

Developing robust statistical tests to detect predictability in economic and financial time series is a central challenge in econometrics. The presence of persistent predictor variables, coupled with correlated and serially dependent shocks affecting both predictors and predictand, creates non-standard testing environments where classical Wald or LM statistics often lead to severely oversized tests. This issue is prevalent in macroeconomic and financial applications, where both highly persistent series (e.g., inflation, interest rates) and noisier series (e.g., growth rates in economic activity, currency and stock returns) are frequently encountered. 
How can we reliably test for predictability under uncertain degrees of persistence and with mutually correlated, serially dependent shocks? This paper proposes a novel approach that addresses these challenges and remains simple to implement.

Much of the recent predictive regression literature has shared similar objectives (see for instance Phillips (2015) for a comprehensive survey of the field). Besides offering an original alternative to these existing methods, we argue that in some instances our setting allows for greater generality, particularly when it comes to accommodating different classes of persistent processes co-existing as predictors, and whose statistical relevance needs to be assessed in a setting where shocks to predictand and predictors are mutually correlated and serially dependent.

Numerous methods for robustifying inferences in predictive regressions have been developed in recent years. A particularly effective and popular technique among practitioners is the instrumental variable approach of Phillips and Magdalinos (2009), commonly referred to as IVX and further formalized in the specific context of predictive regressions as used in the finance literature by Kostakis, Magdalinos, and Stamatogiannis (2015, 2023) and Magdalinos (2022). This method involves assuming that all predictors fall within a particular persistence class ranging from purely stationary to purely integrated, and subsequently using these predictors to generate instruments whose degree of persistence matches that of a nearly-stationary process by design. These instruments are then used to construct a suitable Wald statistic whose asymptotic distribution is shown to be a nuisance-parameter-free $\chi^{2}$
regardless of the persistence class to which the original predictors belong. Another contribution within the same area and with similar objectives is the variable addition-based approach developed by Breitung and Demetrescu (2015), which also relies on generated IVs but is implemented within a suitably augmented predictive regression model.

The method introduced in this paper differs from the above in that it does not require any instrumentation nor any modification/augmentation of the predictive regression model under consideration. The approach is based on a classical formulation of a Wald-type statistic expressed as a suitably standardized difference between the restricted and unrestricted sums of squared residuals. This spread between these two sums of squared residuals is modified by hitting the unrestricted squared residuals with a particular \emph{function} of a Bernoulli sequence. This is done in a way that induces no data loss and results in a favorable change in the underlying asymptotics of the modified Wald-type statistic, rendering it robust to the stochastic properties of predictors and to serial dependence in the errors driving both predictors and predictand (e.g., heteroskedasticity and/or serial correlation) while also maintaining its robustness to endogeneity arising from non-zero contemporaneous correlations between the shocks to the predictand and the predictors. This robustness is achieved without any form of correction or adjustment to the proposed test statistics. Unlike existing methods, the asymptotic behavior of our test statistics is also insensitive to the inclusion of intercept terms in the autoregressive dynamics of the predictors. 

To motivate our approach we first introduce a test statistic based on a single Bernoulli sequence which we subsequently generalize to a formulation designed to accommodate multiple such sequences. The introduction of a Bernoulli sequence in our test statistics introduces a tuning parameter: its success probability. The impact of the choice of this tuning parameter is shown to be inconsequential for the null distributions of the proposed tests provided that this probability is bounded away from 0, 1, and 1/2. The invariance of the null distributions to this tuning parameter in turn allows us to guide its choice via its impact on their local power properties. This approach ensures that the resulting inferences exhibit excellent size control and good power throughout. 

The paper is structured as follows. Section 2 introduces the regression-based testing environment and the novel test statistics with a focus on their underlying motivation. Section 3 obtains their asymptotic distribution under the null hypothesis, and Section 4 establishes their consistency and local power properties. Section 5 focuses on a rigorous theoretical and empirical evaluation of the role of the tuning parameter on these asymptotics. Section 6 provides a comprehensive analysis of their finite-sample properties in settings commonly encountered in applied work, emphasizing their advantages and shortcomings relative to existing methods. Section 7 overviews the practical implementation of the methods and concludes.

\section{Modelling Environment and Test Statistics for Predictability}

\subsection{Modelling Environment}

Given $n$ observations $\{y_{t}\}_{t=1}^{n}$ and $\{{\bm x}_{t}\}_{t=1}^{n}=\{(x_{1t},\ldots,x_{pt})^{\top}\}_{t=1}^{n}$, where $p\geq 1$ ,
the objective is to test the statistical significance of all or a subset of slope parameter estimates obtained from the following predictive regression model
\begin{align}
	y_{t} & = \mu+{{\bm \beta}}^\top {\bm x}_{t-1}+u_{t} \label{eq:eq1} 
\end{align}
where ${\bm \beta}=(\beta_{1},\ldots,\beta_{p})^{\top}$ is a vector of $p$ slope parameters associated with 
the p-dimensional predictor vector ${\bm x}_{t-1}$ and $u_{t}$ is a random disturbance term. For notational convenience, define $\widetilde{{\bm x}}_{t}^{\top}=(1,x_{1t},\ldots,x_{pt})$ and ${\bm \theta}^{\top}=(\mu,{\bm \beta}^{\top})$, and  rewrite equation (\ref{eq:eq1}) as
\begin{align}
y_{t} & ={\bm \theta}^{\top} \widetilde{\bm x}_{t-1}+u_{t}.
 \label{eq:eq2}
\end{align}
The predictor variables are modelled as a vector autoregressive process 
\begin{align}
	{\bm x}_{t} & = {\bm \varphi_{0}}+\left({\bm I}_{p}-{\bm D}_{n}(\bm \alpha)^{-1} \ {\bm C}\right) {\bm x}_{t-1}+{\bm v}_{t} \label{eq:eq3}
\end{align}
\noindent
where ${\bm D}_{n}(\bm \alpha)=\textrm{diag}(n^{\alpha_{1}},\ldots, n^{\alpha_{p}})$, ${\bm C}=\textrm{diag}(c_{1},\ldots,c_{p})$, $c_{i}>0$, and ${\bm v}_{t}$ is a p-dimensional possibly dependent random disturbance vector.
The parameters $\alpha_{i}$ and $c_{i}$ govern the persistence of the predictors. 
 Depending on the magnitudes of these $\alpha_{i}'s$ and $c_{i}'s$ which may or may not be all identical, such a formulation allows for a great degree of flexibility in capturing the potentially persistent nature of predictors, in addition to allowing this persistence to be heterogeneous across different predictors. 

Throughout this paper attention is restricted to three classes of predictors, encompassing the vast majority of time-series data commonly used in empirical research. These three classes are labelled as ${\cal C}_{1}$, ${\cal C}_{2}$ and ${\cal C}_{3}$ and correspond to stationary, nearly-stationary and nearly integrated parameterizations of (\ref{eq:eq3}). Formally, for any predictor $x_{it}$ from the pool of $p$ predictors in ${\bm x}_{t}$:
\begin{itemize}
	\item[(i)] $x_{it} \in {\cal C}_{1}$ if $\alpha_{i}=0$ and $|1-c_{i}|<1$
	\item[(ii)] $x_{it} \in {\cal C}_{2}$ if $\alpha_{i} \in (0,1)$ and $c_{i}>0$
	\item[(iii)] $x_{it} \in {\cal C}_{3}$ if $\alpha_{i}=1$ and $c_{i}>0$
\end{itemize}
The notational convention ${\cal C}_{\ell}(x_{i})={\cal C}_{\ell}(x_{j})$ refers to two predictors $x_{it}$ and $x_{jt}$ that belong to the same class $\ell \in \{1,2,3\}$. \\

\noindent 
{\bf Remark 1.} We note from the specification of the predictors in (\ref{eq:eq3}) that their VAR dynamics are allowed to accommodate non-zero intercepts, regardless of the underlying magnitudes of the $\alpha_{i}'s$. The methods developed in the sequel are unaffected by the presence or absence of such  intercept terms. This is in contrast to existing methods which typically require a zero intercept for the validity of their asymptotics (see Remark 4 in Yang, Liu, Peng and Cai (2021, p. 689)). \\

Given this modelling environment, our objective is to test the null hypothesis $\textrm{H}_{0}\colon \textrm{R} \ {\bm \beta}=0$ against the alternative hypothesis $\textrm{H}_{1}\colon \textrm{R} \ {\bm \beta}\neq 0$ where $\textrm{R}$ denotes a user-specified restriction matrix of dimension $r \times p$ and of rank $r$. The tests are designed to be invariant to the $\alpha_{i}'s$ and $c_{i}'s$, to allow for the presence of conditional heteroskedasticity and serial correlation in the disturbances $u_{t}$ and ${\bm v}_{t}$ and to also accommodate endogeneity in the form of non-zero contemporaneous correlations between $u_{t}$ and the $v_{jt}'s$.   

\subsection{Test Statistics}

To motivate our test statistics, let us first consider the classical linear model setting and recall the standard Wald-type statistic formulation, which compares restricted and unrestricted sums of squared residuals from a least squares regression.
Letting $\hat{u}_{0t}$ and $\hat{u}_{1t}$ denote these residuals, and $\hat{\sigma}^{2}_{0}$ and $\hat{\sigma}^{2}_{1}$ their corresponding variances, recall: 
\begin{align}
\textrm{W}_{n} & = \dfrac{n(\hat{\sigma}^{2}_{0}-\hat{\sigma}^{2}_{1})}{\hat{\sigma}^{2}_{1}} \coloneqq \dfrac{n \ \textrm{N}_{n}}{\hat{\sigma}^{2}_{1}}
	\label{eq:eq4}
\end{align}
which under standard regularity conditions and $x_{it} \in {\cal C}_{1}$ $\forall i \in [p]$ in particular,
satisfies $W_{n}\stackrel{d}\rightarrow \chi^{2}(r)$ under the null hypothesis. 
This classical $\chi^{2}$ limit derives from establishing that $n \ \textrm{N}_{n}=O_{p}(1)$ which upon suitable scaling with $\hat{\sigma}^{2}_{1}$ results in the familiar nuisance-parameter free $\chi^{2}(r)$ limit.  

However, when the fitted model contains predictors with persistent dynamics (i.e., belonging to classes ${\cal C}_{2}$ or ${\cal C}_{3}$) and when there are substantial correlations between the errors of the predictors and the predictand, the chi-squared asymptotic approximation becomes unreliable, often leading to oversized tests. This has motivated the development of the predictive regression literature, which aims to explicitly model persistence as in equation (3), with the expectation that the resulting non-standard asymptotics will provide more accurate approximations for testing hypotheses such as $H_{0}\colon R{\bm \beta}=0$. The caveat with this approach is that these non-standard asymptotics depend on the parameters used to model persistence (e.g., the $c_{i}'s$ and $\alpha_{i}'s$ in equation (\ref{eq:eq3})). Recent literature, such as the IVX methodology, has sought to address this issue by introducing alternative estimation methods designed to make inferences robust to the precise nature of the persistence present in the model. In what follows we motivate and introduce an alternative to such existing methods

Consider a modification of the numerator of (\ref{eq:eq4}). Let 
$\{b_{t}\}_{t=1}^{n}$ denote an i.i.d Bernoulli sequence that is independent of the data and with a given probability of success $p_{0}$ ($b_{t}=1$ with probability $p_{0}$ and $b_{t}=0$ with probability $(1-p_{0})$ for $0<p_{0}<1$), and consider
\begin{align}
	\widetilde{\textrm{N}}_{n}(p_{0}) & = \frac{1}{2}\left(
	\frac{\sum_{t=1}^{n}b_{t}\hat{u}_{0t}^{2}}{\sum_{t=1}^{n} b_{t}}
	+\frac{\sum_{t=1}^{n}(1-b_{t})\hat{u}_{0t}^{2}}{\sum_{t=1}^{n} (1-b_{t})}
	\right)-\frac{1}{n} \sum_{t=1}^{n}\hat{u}_{1t}^{2} \label{eq:eq5} \\
& =	\frac{1}{2}\left(
	\frac{\sum_{t=1}^{n}b_{t}\hat{u}_{0t}^{2}}{n \overline{b}_{n}}
	+\frac{\sum_{t=1}^{n}(1-b_{t})\hat{u}_{0t}^{2}}{n(1-\overline{b}_{n})}
	\right)-\frac{1}{n} \sum_{t=1}^{n}\hat{u}_{1t}^{2} \label{eq:eq6} \\
& = \frac{1}{n} \sum_{t=1}^{n} w_{t}(p_{0})\hat{u}_{0t}^{2}- \frac{1}{n} \sum_{t=1}^{n}\hat{u}_{1t}^{2} \label{eq:eq7} \\
& = \frac{1}{n} \sum_{t=1}^{n} w_{t}(p_{0})(\hat{u}_{0t}^{2}-\hat{\sigma}^{2}_{1})- \frac{1}{n} \sum_{t=1}^{n}(\hat{u}_{1t}^{2}-\hat{\sigma}^{2}_{1}) \label{eq:eq8} \\
& \coloneqq \frac{1}{n} \sum_{t=1}^{n} d_{t}(p_{0})	
\label{eq:eq9}
\end{align}
\noindent
where 
\begin{align}
	w_{t}(p_{0}) & \coloneqq \frac{1}{2}\left(\dfrac{b_{t}}{\overline{b}_{n}}+\dfrac{1-b_{t}}{(1-\overline{b}_{n})}\right)
	\label{eq:eq10}
\end{align}
and $\hat{\sigma}^{2}_{1}$ in (\ref{eq:eq8}) denotes a generic consistent estimator of $E[u_{t}^{2}]$ formed using $\hat{u}_{1t}^{2}$. Note from 
(\ref{eq:eq10}) that $\sum_{t=1}^{n}w_{t}(p_{0})=n$ $\forall p_{0}$ by construction so that (\ref{eq:eq8}) is just a reformulation of (\ref{eq:eq7}) which will be convenient when establishing the asymptotics properties of the tests. For later use, define $w_{t}^{0}(p_{0})$ as:
\begin{align}
	w_{t}^{0}(p_{0}) & \coloneqq \frac{1}{2}\left(\dfrac{b_{t}}{p_{0}}+\dfrac{1-b_{t}}{(1-p_{0})}\right).
	\label{eq:eq11}
\end{align}
which replaces the sample means $\overline{b}_{n}$ with their population counterparts in $w_{t}(p_{0})$. \\

Before proceeding further it is useful to highlight some key characteristics of the $w_{t}(p_{0})$ sequence in (\ref{eq:eq10}) that will play an important role in the sequel. Note for instance that $w_{t}(p_{0})$ is not i.i.d due to the presence of the entire Bernoulli sequence appearing in its denominator via $\overline{b}_{n}$. It will however share the same characteristics (e.g., moments) as the i.i.d. sequence $w_{t}^{0}(p)$ asymptotically. Invoking the strong law of large numbers (SLLN) combined with the continuous mapping theorem it is also straightforward to establish that the sample moments and autocovariances of $w_{t}(p_{0})$ converge almost surely to the population moments of $w_{t}^{0}(p_{0})$. Focusing on the expectation and variance of $w_{t}(p_{0})$, the following convergence of moments hold as $n \rightarrow \infty$
\begin{align}
E_{b}[w_{t}(p_{0})] & \rightarrow E_{b}[w_{t}^{0}(p_{0})]=1 \ \ \forall p_{0}	\label{eq:eq12} \\
V_{b}[w_{t}(p_{0})] & \rightarrow V_{b}[w_{t}^{0}(p_{0})]=\frac{(1-2p_{0})^{2}}{4p_{0}(1-p_{0})}.
	\label{eq:eq13}
\end{align}

\noindent
Noting that under $p_{0}=1/2$, $V_{b}[w_{t}^{0}(p_{0}=1/2)] = 0$ which in turn implies that $w_{t}^{0}(p_{0}=1/2)=1$ it directly follows that 
\begin{align}
	E_{b}\ \left[\dfrac{n \ \widetilde{\textrm{N}}_{n}(p_{0}=1/2)}{\hat{\sigma}^{2}_{1}}\right] & \approx  \dfrac{n \ \textrm{N}_{n}}{\hat{\sigma}_{1}^{2}}
		\label{eq:eq14}
\end{align}
so that $\widetilde{\textrm{N}}_{n}(p_{0})$ in (\ref{eq:eq5})-(\ref{eq:eq9}) can be viewed as a generalized version of $\textrm{N}_{n}$ in the classical Wald statistic formulation in the sense of specializing to it in expectation  when $p_{0}=1/2$. \\

Our initial proposal is to consider inferences based on  $\widetilde{\textrm{N}}_{n}(p_{0})$ as formulated in (\ref{eq:eq9}) with a single $b_{t}$ sequence having its associated probability of success bounded away from zero, one, and one-half to ensure a well defined and non-degenerate variance function in (\ref{eq:eq13}). By doing so the asymptotic behaviour of the numerator of  $\textrm{W}_{n}$ changes fundamentally in the sense that $\sqrt{n} \ \widetilde{\textrm{N}}_{n}(p_{0})$ becomes $O_{p}(1)$ instead of its classical $n \ \textrm{N}_{n}=O_{p}(1)$ behaviour. As we establish further below,  
when suitably standardized by its variance the resulting distribution of $\sqrt{n}\ \widetilde{\textrm{N}}_{n}(p_{0})$ becomes
invariant to the properties of predictors and other problematic features (e.g., heteroskedasticity and serial correlation) {\it provided} that 
the squared residuals are estimated consistently. \\

\noindent
{\bf Remark 2}. The formulation of $\widetilde{N}_{n}(p_{0})$ in (\ref{eq:eq5}) can be viewed as replacing the standard sample average  $\hat{\sigma}^{2}_{0}\coloneqq \sum_{t=1}^{n}\hat{u}_{0t}^{2}/n$ with a split-sample average. Specifically, the unrestricted squared residuals  
$\hat{u}_{0t}^{2}$ are randomly assigned to two groups, and their group means are subsequently averaged as $0.5(\sum b_{t}\hat{u}_{0t}^{2}/\sum b_{t} + \sum (1-b_{t})\hat{u}_{0t}^{2}/\sum (1-b_{t}))$. This  highlights the important role played by the specific choice of $p_{0}=1/2$, under which the above split-sample average becomes equivalent to 
$\sum_{t=1}^{n}\hat{u}_{0t}^{2}/n$ for large enough $n$ but is otherwise different in the sense of having a mildly different variance which can be controlled via $p_{0}$
and made arbitrarily close to that of $\hat{\sigma}^{2}_{0}$.   \\

The first test statistic denoted ${\cal S}_{n}(p_{0})$ is now formulated as: 
\begin{align}
	{\cal S}_{n}(p_{0}) & = \left(\dfrac{\sqrt{n} \ \overline{d}_{n}}{s_{d}}\right)^{2}
	\label{eq:eq15}
\end{align}
where $\overline{d}_{n}=\sum_{t=1}^{n}d_{t}(p_{0})/n$ for $d_{t}(p_{0})=w_{t}(p_{0})(\hat{u}_{0t}^{2}-\hat{\sigma}^{2}_{1})-(\hat{u}_{1t}^{2}-\hat{\sigma}^{2}_{1})$ as in (\ref{eq:eq8}) and $s^{2}_{d}=n^{-1}\sum_{t=1}^{n}(d_{t}(p_{0})-\overline{d}_{n})^{2}$. This is a is a classical studentized sample mean (squared) expressed as a function of a given tuning parameter $p_{0}$. \\

{\bf Remark 3}. To illustrate the advantages of using a test statistic based on $\widetilde{\textrm{N}}_{n}(p_{0})$ defined in (\ref{eq:eq9}) and leading to (\ref{eq:eq15}), with a suitable $p_{0}$, consider the simple model $y_{t} = \beta x_{t-1} + u_{t}$. Under the null hypothesis we have $\sum_{t=1}^{n} \hat{u}_{0t}^{2}=\sum_{t=1}^{n}u_{t}^{2}$
and under the alternative $\sum_{t=1}^{n} \hat{u}_{1t}^{2}=\sum_{t=1}^{n}u_{t}^{2}-(\sum x_{t-1}u_{t}/\sqrt{\sum x_{t-1}^{2}})^{2}$.
Thus, $\textrm{N}_{n} = n^{-1} \left( \sum x_{t-1} u_{t} / \sqrt{\sum x_{t-1}^{2}} \right)^{2}$, and $n \ \textrm{N}_{n} = O_{p}(1)$ under standard regularity conditions. Using $\widetilde{\textrm{N}}_{n}(p_{0})$ instead, results in 
$n^{-1/2} \ \widetilde{\textrm{N}}_{n}(p_{0}) = \sqrt{n} \ \sum_{t=1}^{n}(w_{t}^{0}(p_{0})-1)(u_{t}^{2}-\sigma^{2}_{u})+o_{p}(1)$ 
so that the null asymptotics become driven by $n^{-1/2} \sum_{t=1}^{n} (w_{t}^{0}(p_{0}) - 1) (u_{t}^{2} - \sigma_{u}^{2})$, which, under very mild assumptions that do not need to restrict the dependence of the $u_{t}'s$, results in nuisance parameter free distributions which also remain robust to the stochastic properties of predictors. These observations also demonstrate how the proposed approach breaks-up the potential presence of dependence in $(u_{t}^{2}-\sigma^{2}_{u})$ due to the martingale difference structure of 
$(w_{t}^{0}(p_{0})-1)(u_{t}^{2}-\sigma^{2}_{u})$ (this regardless of the dependence structure of $(u_{t}^{2}-\sigma^{2}_{u}))$. Inferences based on ${\cal S}_{n}(p_{0})$  will solely require an estimator of $E[(w_{t}^{0}(p_{0})-1)^{2}(u_{t}^{2}-\sigma^{2}_{u})^{2}]$, thus bypassing the need to appeal to HAC type normalizers. Crucially, this will continue to be the case regardless of whether $x_{t}$ is persistent or not. \\

Although the above analysis implies that we will be able to ignore the presence of conditional heteroskedasticity and base inferences on standard homoskedastic variance estimates this will not {\it always} be the case if we also wish to allow the $u_{t}'s$ to be serially correlated. The key reason for this is the need for 
the squared residuals to remain consistent for their true counterparts under the null hypothesis. This cannot hold for instance if $x_{t} \in {\cal C}_{1}$ (i.e., a purely stationary sequence), and endogeneity (e.g., $E[u_{t}x_{t}]\neq 0$) is combined with serial correlation in the $u_{t}'s$ since $\hat{\beta}$ would no longer be consistent for $\beta$, a standard result in regression models estimated via least-squares. Nevertheless, there will be multiple empirically relevant instances where serial correlation in the $u_{t}'s$ will also not be of any concern for the validity of our inferences as for instance when $x_{t}$ is mildly or highly persistent (regardless of whether endogeneity is present or not) or $x_{t}$ is purely stationary but endogeneity is ruled out. It is well known for instance that when $x_{t}$ is nearly integrated $\hat{\beta}$ remains consistent for $\beta$ regardless of whether the $u_{t}'s$ are serially correlated  or not. \\

The practical implementation of ${\cal S}_{n}(p_{0})$ necessitates a user-specified parameter, the implications of which are thoroughly examined in Section 5, following 
the formal local power analysis of ${\cal S}_{n}(p_{0})$ developed in Section 4.   The key message conveyed both theoretically and empirically is that the choice of $p_{0}$, provided that it is bounded away from 0, 1,  and 1/2, has no theoretical impact on the null asymptotics 
and this robustness is adequately maintained in finite samples as well.  This will lead us to 
suggest setting $p_{0}$ in a way that maximizes local power, a strategy shown to be highly effective in the sequel. \\ 

Given that the proposed test statistic relies on an external Bernoulli sequence via $w_{t}(p_{0})$ one may conjecture that implementing ${\cal S}_{n}(p_{0})$ across multiple such sequences may enhance power through the aggregation of potentially weak signals. To this end,   
we introduce the following generalization of the earlier {\it single-shot statistic}
\begin{align}
	{\cal S}_{\textrm{M}}(p_{0}) & = \sum_{j=1}^{\textrm{M}} {\cal S}_{n,j}(p_{0}) \label{eq:eq16}
\end{align}
and which can be seen to specialize to ${\cal S}_{n}(p_{0})$ for $\textrm{M}=1$. 
Here, the ${\cal S}_{j,n}(p_{0})'s$ are given by (\ref{eq:eq15}) across \textrm{M} mutually independent, and i.i.d. Bernoulli 
sequences $\{b_{jt}\}_{j=1}^{\textrm{M}}$ with the same probability of sucess $p_{0}$. As we establish in the sequel, this test statistic has a $\chi^{2}(\textrm{M})$ distribution under the null hypothesis, so that if we wish to operate under a large $\textrm{M}$ that is potentially growing with $n$, we can concentrate on its centered and scaled version 
\begin{align}
	{\cal Q}_{M_{n}}(p_{0}) & = \frac{{\cal S}_{M_{n}}(p_{0})-M_{n}}{\sqrt{2M_{n}}}. \label{eq:eq17}
\end{align}
Basing inferences on (\ref{eq:eq17}) would allow us to bypass the need to specify an exact magnitude for $\textrm{M}$ in ${\cal S}_{M}(p_{0})$, provided that the asymptotic behaviour of ${\cal Q}_{M_{n}}(p_{0})$ as $(n,M_{n})\rightarrow \infty$ can guide us on suitable growth rates for $\textrm{M}_{n}$ relative to $n$, as it is commonly done in contexts such as bandwidth selection.

\section{Null Asymptotics}

The focus here is on obtaining the limiting distributions of ${\cal S}_{\textrm{M}}(p_{0})$ and ${\cal Q}_{M_{n}}(p_{0})$ when one wishes to test the null hypothesis $H_{0}\colon \textrm{R} \ {\bm \beta}=0$ within (\ref{eq:eq1}).  Given the multitude of scenarios arising from the various classes of persistence and their combinations, and  to highlight the robustness of the proposed approach to these, we choose to partly operate under a set of high level assumptions which rely on well known results in the literature. These operating assumptions are collected in the sets labelled as {\bf A} and {\bf B} below. \\

\noindent
{\bf Assumptions A:} ({\bf A1}) {\it $\{b_{j,t}\}_{t=1}^{n}$ for $j=1,\ldots,M$ are $\textrm{M}$ i.i.d. Bernoulli sequences that are mutually independent and independent of the data, each having the same probability of success $p_{0} \in {\cal P}_{\epsilon_{a},\epsilon_{b}}$ for ${\cal P}_{\epsilon_{a},\epsilon_{b}}=\{p \colon \ 0<\epsilon_{a} \leq p \leq 1-\epsilon_{a}<1, |p-\frac{1}{2}| \ \geq \epsilon_{b}\}$, for some positive constants $\epsilon_{a}$ and $\epsilon_{b}$. ({\bf A2}) Let $\eta_{t} \coloneqq u_{t}^{2}-E[u_{t}^{2}]$. The centered sequence $\{\eta_{t}\}_{t=1}^{n}$ is strictly stationary and ergodic with $E|\eta_{t}|^{2+\kappa}<\infty$ for some $\kappa>0$. We write its variance as $\sigma^{2}_{\eta}\coloneqq E[\eta_{t}^{2}]$.  ({\bf A3}) A consistent estimator $\hat{\sigma}^{2}_{\eta}$ of $E[\eta_{t}^{2}]$ exists, that is
$\hat{\sigma}^{2}_{\eta} \stackrel{p}\rightarrow \sigma^{2}_{\eta}$.} \\

\noindent
{\bf Assumptions B:} {\it With ${\bm K}_{n}=diag(n^{\frac{1}{2}},n^{\frac{1+\alpha_{1}}{2}},\ldots,n^{\frac{1+\alpha_{p}}{2}})$ and as $n\rightarrow \infty$ it holds that ({\bf B1})  ${\bm K}_{n}^{-1}\sum_{t=1}^{n} \widetilde{\bm x}_{t-1}u_{t} \stackrel{d}\longrightarrow {\bm G}_{\infty}$, a centered p-dimensional $O_{p}(1)$ random vector process, and ({\bf B2}) ${\bm K}_{n}^{-1}\sum \widetilde{\bm x}_{t-1}\widetilde{\bm x}_{t-1}^{\top}{\bm K}_{n}^{-1}=O_{p}(1)$.} \\

Assumption {\bf A1} introduces the external process with which the allocation of unrestricted residuals is randomized when computing their split-sample average as in (\ref{eq:eq5}). The main requirement is for $p_{0}$ to be bounded away from zero, one and one-half. Specifically $\epsilon_{a}$ and $\epsilon_{b}$ refer to the tolerance levels associated with these three boundaries. 
Assumption {\bf A2} ensures that a suitable CLT for a particular multiplicative process arising from 
the formulation in (\ref{eq:eq8})-(\ref{eq:eq9}) can be invoked. It imposes a mild structure on the dependence of $\eta_{t}$ without restricting it to be an m.d.s for instance. It may for example, take the form of a strong mixing process. The particular moment restriction is made for the purpose of verifying the validity of the conditional variance and Lindeberg conditions when 
invoking a suitable CLT. Importantly, no assumptions are imposed on the long-run variance of the $\eta_{t}'s$, requiring solely the availability of a consistent estimator of its naive variance via {\bf A3}. \\

Assumptions {\bf B} ensure that the model parameters and the squared residuals can be estimated consistently. They also ensure that the least-squares based estimators of $\mu$ and ${\bm \beta}$ have well defined limiting distributions upon applying suitable normalizations. Note that there is no requirement for these limiting distributions to be free of nuisance parameters. An immediate implication of {\bf B1}-{\bf B2} for instance is that $\sqrt{n}(\hat{\mu}-\mu)=O_{p}(1)$ and $n^{\frac{1+\alpha_{i}}{2}}(\hat{\beta}_{i}-\beta_{i})=O_{p}(1)$.  Examples of primitive conditions ensuring that {\bf B1}-{\bf B2} hold can be found in Phillips and Magdalinos (2009), Kostakis, Magdalinos and Stamatogiannis (2015), Magdalinos (2023) among others. Note that in {\bf B2} the formulation of the limit of the 
normalized sample moment matrix is left unspecified as its explicit expression will vary depending on the type of persistence under consideration. The key point is that under the null hypothesis, the nature of these limits and hence the stochastic properties of the predictors will not matter beyond requiring that these limits  are finite and well-defined.   \\
 
Assumptions {\bf B} accommodate contemporaneous endogeneity (e.g., $E[x_{it}u_{t}]\neq 0$), conditional heteroskedasticity and serial correlation in the $u_{t}'s$ and $v_{t}'s$ as long as consistent estimation of $\mu$ and ${\bm \beta}$ is possible. It is well known that under persistent covariates (i.e., if all $\alpha_{i}'s$ are in $(0,1]$) conditional heteroskedasticity and/or serial-correlation do not asymptotically affect the consistency of the least-squares estimators regardless of whether endogeneity is present or not. Intuitively, when covariates are persistent the strength of their signal eliminates the impact of such effects on estimator consistency. Interestingly consistency continues to hold under conditional heteroskedasticity even for $\alpha_{i}=0$ (i.e., pure stationarity), resulting in our proposed inferences being robust to conditional heteroskedasticity for any $\alpha \in [0,1]$ without requiring any adjustment to the test statistics. Although the above assumptions also naturally accommodate serial correlation if all $\alpha_{i}'s$ are in $(0,1]$, this will not be the case under $\alpha_{i}=0$ if endogeneity is present as in such instances consistent estimation would not be possible.   \\

The following proposition states our first result about the asymptotic behaviour of ${\cal S}_{\textrm{M}}(p_{0})$ under the null hypothesis of interest and any fixed integer magnitude $\textrm{M}$. \\

\noindent
{\bf Proposition 1.} \emph{Under the null hypothesis $\textrm{H}_{0}\colon \textrm{R} \ {\bm \beta}=0$, and assumptions {\bf A1}-{\bf A3}, and {\bf B1}-{\bf B2} we have 
	\begin{align}
		{\cal S}_{\textrm{M}}(p_{0}) & \stackrel{d}\rightarrow \chi^{2}(\textrm{M}).
			\label{eq:eq18}
	\end{align} 
	as $n \rightarrow \infty$}. \\

The result in (\ref{eq:eq18}) shows that, for any fixed $\textrm{M}$, the limiting distribution is $\chi^{2}(\textrm{M})$ {\it irrespective} of the underlying persistence of predictors and dependence structure of the random disturbances. 
Next, we consider the case of the test-statistic ${\cal Q}_{M_{n}}(p_{0})$. \\

\noindent
{\bf Proposition 2.} \emph{Suppose $\textrm{M}_{n}\rightarrow \infty$ with $\textrm{M}_{n}/n \rightarrow 0$ as $n \rightarrow \infty$. Under the null hypothesis $\textrm{H}_{0}\colon \textrm{R} \ {\bm \beta}=0$, and assumptions {\bf A1}-{\bf A3}, and {\bf B1}-{\bf B2} we have 
	\begin{align}
		{\cal Q}_{\textrm{M}_{n}}(p_{0}) & \stackrel{d}\rightarrow {\cal N}(0,1).
		\label{eq:eq19}
	\end{align} 
}. \\

An important take away from our results in Propositions 1 and 2 is that under the null hypothesis the asymptotic distributions of both test statistics are invariant to $p_{0}$ provided that $p_{0} \in {\cal P}_{\epsilon_{a},\epsilon_{b}}$. As documented 
in the sequel, this robustness of the null asymptotics to $p_{0} \in {\cal P}_{\epsilon_{a},\epsilon_{b}}$ also extends to finite samples 
with these test statistics shown to maintain excellent size control across different magnitudes of $p_{0}$. 
Given the expression of the variance of $w_{t}(p_{0})$ in (\ref{eq:eq13}) 
we naturally expect distortions to kick in as $p_{0}$ approaches the variance degeneracy boundary of 1/2, causing the test statistics to concentrate more and more tightly around zero. Our results in Propositions 1-2 are not designed to be informative about such phenomena beyond saying that for sufficiently large $n$ choices of $p_{0}$ do not matter for the $\chi^{2}$ and Gaussian approximations. We postpone a more in-depth analysis of the role of $p_{0}$ until after the local power analysis that follows.

\section{Local Power}

We now consider the limiting properties of the test statistics under local alternatives. We concentrate on testing the global null hypothesis $H_{0} \colon {\bm \beta} = {\bm 0}$ against 
alternatives parameterized locally to this global null. Given the specifications in (\ref{eq:eq1})-(\ref{eq:eq2}) we formulate these local departures from the null as
\begin{align}
\textrm{H}_{1n} \colon & \ \ {\bm \beta}^{\top}_{n} = \left(\dfrac{\delta_{1}}{n^{\frac{1+2\alpha_{1}}{4}}},\ldots,
\dfrac{\delta_{p}}{n^{\frac{1+2\alpha_{p}}{4}}}\right) \coloneqq  n^{\frac{1}{4}} \ 
(\overline{\bm K}_{n}^{-1} \ {\bm \delta})^{\top}
\label{eq:eq20}
\end{align}  
for $\overline{\bm K}_{n}=diag(n^{\frac{1+\alpha_{1}}{2}},\ldots,n^{\frac{1+\alpha_{p}}{2}})$. 
The true model is understood to be given by (\ref{eq:eq1}) with the $\beta_{i}'s$ as in (\ref{eq:eq20}) and where all, some or just one of the $\delta_{i}'s$ are non-zero. For notational simplicity and no loss of generality we also set $\mu=0$ throughout while the fitted models are always understood to contain fitted intercepts. \\

We initially restrict the local power analysis to a scenario with predictor homogeneity in the sense that all $p$ predictors are understood to belong to one of the three persistence classes of interest. We write this as ${\cal C}_{\ell}(x_{1})=\ldots = {\cal C}_{\ell}(x_{p})$ for $\ell \in \{1,2,3\}$. Note that this does not necessarily require all 
$\alpha_{i}'s$ to be identical in the sense that under the specific class ${\cal C}_{2}$ the $\alpha_{i}'s$ can take distinct values within $(0,1)$. \\

Unlike the earlier null asymptotics the type of persistence characterising the $p$ predictors will now have a direct influence on the local power properties of the tests. This can be seen from the formulation of $H_{1n}$ in (\ref{eq:eq20}) where the rates of departure from $H_{0}$ are determined by the magnitudes of the $\alpha_{i}'s$. The closer the $\alpha_{i}'s$ are to one, the closer to $H_{0}$ the local alternatives $H_{1n}$ are allowed to be. Specializing (\ref{eq:eq20}) to a single predictor scenario with slope parameter $\beta_{1}$ we can observe for instance than in a purely stationary context that corresponds to $\alpha_{1}=0$, the local 
departures are parameterized as $\beta_{1}=\delta_{1}/n^{\frac{1}{4}}$ while under 
near-integratedness which corresponds to $\alpha_{1}=1$ we have 
$\beta_{1}=\delta_{1}/n^{\frac{3}{4}}$. These differences parallel the influence of predictor persistence on the signal to noise ratios (SNR) of the underlying regression model in (\ref{eq:eq2})-(\ref{eq:eq3}). Holding the noise variance constant, the greater this persistence, the higher the SNR will be.  \\

\noindent 
{\bf Remark 4.} From the parameterization of the departures from the null hypothesis in (\ref{eq:eq19}) one may speculate that it is the reliance of our test statistics on the split-sample averaging device that is inducing the need to consider more pronounced departures from the null relative to classical testing environments (e.g., $n^{-1/4}$ versus $n^{-1/2}$ in a stationary setting and $n^{-3/4}$ versus $n^{-1}$ in a nearly-integrated setting). 
This however is by no means related to the particular randomized averaging process used to construct ${\cal S}_{M}(p_{0})$ and would in fact occur under any testing environment in which inferences are based on squared errors rather than their levels, and consequently on the $\beta_{i}^{2}$ rather than the $\beta_{i}'s$. Nevertheless, these local parameterization rates do point at the fact that in stationary predictor settings we cannot expect a statistic such as ${\cal S}_{M}(p_{0})$ to power-dominate a test statistic that can accommodate $n^{-1/2}$ as opposed to $n^{-1/4}$-departures from the null (i.e., local-power as opposed to non-local power). Whether these asymptotic based differences translate into equally significant finite sample based power spreads across alternative test statistics is an issue we explore in-depth through simulations.

\subsection{Asymptotic Local Power of ${\cal S}_{M}(p_{0})$ and ${\cal Q}_{M_{n}}(p_{0})$}

We now proceed with the derivation of the asymptotic behavior of ${\cal S}_{M}(p_{0})$ and ${\cal Q}_{M_{n}}(p_{0})$ under $H_{1n}$. For this purpose we start by providing a more explicit characterization of our earlier assumption {\bf B2} depending on whether the $p$ predictors belong to ${\cal C}_{1}$, ${\cal C}_{2}$ or ${\cal C}_{3}$. The normalizing matrix $\overline{\bm K}_{n}$ in (\ref{eq:eq20}) is now understood to specialize to $\overline{\bm K}_{n}=\sqrt{n} \ I_{p}$ under ${\cal C}_{1}$, to 
$\overline{\bm K}_{n}=diag(n^{\frac{1+\alpha_{1}}{2}},\ldots,n^{\frac{1+\alpha_{p}}{2}})$ under ${\cal C}_{2}$ and to $\overline{\bm K}_{n}=n \ I_{p}$ under ${\cal C}_{3}$. Our earlier assumption {\bf B2} is now replaced with {\bf B2(i)}-{\bf B2(iii)} below. \\

\noindent
{\bf Assumptions}: {\bf B2(i)} If $x_{it} \in {\cal C}_{1}$ $\forall i \in [p]$, $\overline{\bm K}_{n}^{-1} \sum {\bm x}_{t-1}{\bm x}_{t-1}^{\top} \overline{\bm K}_{n}^{-1} \stackrel{p}\rightarrow E[{\bm x}_{t-1}{\bm x}_{t-1}^{\top}] \coloneqq {\bm Q}_{1,\infty}$ for ${\bm Q}_{1,\infty}$ a positive definite matrix; {\bf B2(ii)} If $x_{it} \in {\cal C}_{2}$ $\forall i \in [p]$, $\overline{\bm K}_{n}^{-1} \sum {\bm x}_{t-1}{\bm x}_{t-1}^{\top} \overline{\bm K}_{n}^{-1} \stackrel{p}\rightarrow {\bm C}^{-1/2}{\bm \Omega}_{vv}{\bm C}^{-1/2} \coloneqq {\bm Q}_{2,\infty}$ for ${\bm Q}_{2,\infty}$ a positive definite matrix and ${\bm \Omega}_{vv}$ the covariance matrix of the ${\bm v}_{jt}'s$; {\bf B2(iii)} If $x_{it} \in {\cal C}_{3}$ $\forall i \in [p]$, $\overline{\bm K}_{n}^{-1} \sum {\bm x}_{t-1}{\bm x}_{t-1}^{\top} \overline{\bm K}_{n}^{-1} \stackrel{d}\rightarrow {\bm V} \coloneqq {\bm Q}_{3,\infty}$. \\

\noindent
Assumption {\bf B2(i)} corresponds to a standard law of large numbers type of result ensuring that the sample second moments of the predictors and their cross-products accurately approximate their population counterparts. Assumption {\bf B2(ii)} relates to the scenario where all $p$ predictors are moderately persistent so that a law of large numbers with a suitable normalization continues to apply. Assumption {\bf B2(iii)} corresponds to a scenario where all predictors under consideration are modelled as nearly integrated processes, setting $(\alpha_{1},\ldots,\alpha_{p})=(1,\ldots,1)$ in (\ref{eq:eq3}). Here ${\bm Q}_{3,\infty}$ is a matrix of random variables that would typically take the form of stochastic integrals involving Ornstein-Uhlenbeck processes.\\

The local asymptotic power of our tests is now summarised in Proposition 3 below. \\

\noindent
{\bf Proposition 3.} \emph{Suppose that assumptions {\bf A1}-{\bf A3}, assumption {\bf B1}, and either {\bf B2(i)}, {\bf B2(ii)} or {\bf B2(iii)} hold. {\bf (a)} Under the alternative hypothesis $H_{1n}$, and $\textrm{M}$ fixed, we have}
		\begin{align}
		{\cal S}_{\textrm{M}}(p_{0}) & \stackrel{d}\rightarrow \sum_{j=1}^{M} \left[
		{\cal Z}_{j}+
		\dfrac{\sqrt{4 p_{0}(1-p_{0})}}{|1-2p_{0}|} \		
		\frac{{\bm \delta}^{\top} {\bm Q}_{\infty} {\bm \delta}}{\sigma_{\eta}}		
		\right]^{2}
		\label{eq:eq21}
	\end{align}
\emph{where $({\cal Z}_{1},\ldots,{\cal Z}_{M})$ denote \textrm{M} mutually independent standard normal random variables and ${\bm Q}_{\infty} \in \{{\bm Q}_{1,\infty},{\bm Q}_{2,\infty},{\bm Q}_{3,\infty}\}$ under {\bf B2(i)}, {\bf B2(ii)} and {\bf B2(iii)} respectively}. \emph{{\bf (b)} Under the alternative hypothesis $H_{1n}$, and $(n,M_{n})\rightarrow \infty$ such that $M_{n}/n \rightarrow 0$, we have}
	\begin{align}
	{\cal Q}_{M_{n}}(p_{0}) & \stackrel{d}\rightarrow {\cal N}(0,1)+\frac{p_{0}(1-p_{0})}{(1-2p_{0})^{2}} \left(\frac{\bm \delta^{\top} {\bm Q}_{\infty}\bm \delta}{\sigma_{\eta}}\right)^{2}
		\label{eq:eq22}
	\end{align} 
\vspace{0.2cm}

The results in (\ref{eq:eq21})-(\ref{eq:eq22}) illustrate the consistency of the proposed tests and their asymptotic local power properties. We note for instance that power is increasing monotonically with $\|{\bm \delta}\|^{2}$. They also makes explicit the influence on power of the tuning parameter $p_{0}$. Specifically, all else being equal, the closer $p_{0}$ is to $1/2$, the higher power is expected to be. In sufficiently large samples, the null distributions of these test statistics are invariant to the values of $p_{0}$, suggesting that from an asymptotic perspective,  it is optimal to choose a value for $p_{0}$ near $1/2$. Given that assumption {\bf A1} requires $|p_{0}-1/2|\geq \epsilon_{b}$ for a tolerance level $\epsilon_{b}$ (e.g., $\epsilon_{b}=0.10$), a natural choice would be $p_{0}=1/2-\epsilon_{b}$.  \\

\noindent 
{\bf Remark 5.} The result in (\ref{eq:eq21}) also highlights the role of $p_{0}$ through the function
\begin{align}
f(p_{0}) & \coloneqq \frac{4p_{0}(1-p_{0})}{(1-2p_{0})^{2}}
\label{eq:eq23}
\end{align}
which acts as a scaling factor in the noncentrality parameters of the limiting distributions. Although $f(p_{0})$ is not convex over the entire domain ${\cal P}_{\epsilon_{a},\epsilon_{b}}$, it is strictly convex over the two intervals $I_{1}=[\epsilon_{a},1/2-\epsilon_{b}]$ and $I_{2}=[1/2+\epsilon_{b},1-\epsilon_{a}]$. 
Consequently, for $p_{0} \in I_{1}$, the function reaches its maximum at the boundary of the parameter space, specifically at $p_{0}=1/2-\epsilon_{b}$. This characteristic of $f(p_{0})$ motivates and justifies our choice of implementing inferences based on ${\cal S}_{M}(p_{0})$ and ${\cal Q}_{M_{n}}(p_{0})$ with Bernoulli sequences sharing the {\it same} probability of success $p_{0}$, leading to 
a noncentrality component that increases linearly with $\textrm{M}$. The strict convexity of $f(p_{0})$ over $I_{1}$ (or $I_{2}$) implies that $\textrm{M} \times f(p_{0})$, with $p_{0}$ set at the tolerated boundary, will dominate a scenario where different $p_{0j}'s$ are used, resulting in the factor $\sum_{j=1}^{M}f(p_{0j})$ instead of $M\times f(p_{0})$. \\

Another important aspect of the power properties of ${\cal S}_{M}(p_{0})$ that can be inferred from the above analysis concerns the role of $\textrm{M}$. To illustrate, consider a purely stationary setting with $\alpha_{i}=0$, where (\ref{eq:eq21}) specializes to a noncentral $\chi^{2}$ distribution with $\textrm{M}$ degrees of freedom and noncentrality parameter 
\begin{align}
\textrm{ncp}_{\textrm{M}} & =\textrm{M} \ \underbrace{\frac{4p_{0}(1-p_{0})}{(1-2p_{0})^{2}} \ \left(\frac{\bm \delta' \bm Q_{\infty} \bm \delta}{\sigma_{\eta}}\right)^{2}}_{\bm \lambda}.
\label{eq:eq24}
\end{align}
This highlights the fact that power can be enhanced by adjusting $\textrm{M}$. However, diminishing returns are also expected to kick in as the increase in the critical value due to the increase in the degrees of freedom will at some point outpace the gain in power from the larger noncentrality parameter. Nevertheless, the key point here is that $\textrm{M}$ can be leveraged to offset the less favorable departures from the null hypothesis of order $n^{1/4}$ when predictors are purely stationary or weakly persistent. These points, along with specific guidelines for selecting $\textrm{M}$, are discussed in greater detail in Section 5. \\

Next, we consider a scenario with a mixture of predictor types. Suppose that the $p$ predictors are composed of $p_{1}$ predictors belonging to ${\cal C}_{1}$, $(p_{2}-p_{1})$ predictors belonging to ${\cal C}_{2}$ and $(p-p_{2})$ predictors belonging to ${\cal C}_{3}$. We further let ${\bm \delta}_{1}=(\delta_{1},\ldots,\delta_{p_{1}})$, ${\bm \delta}_{2}=(\delta_{p_{1}+1},\ldots,\delta_{p_{2}})$
and ${\bm \delta}_{3}=(\delta_{p_{2}+1},\ldots,\delta_{p})$. The asymptotic power under this mixed-predictor scenario is summarized in Proposition 4 below. \\

\noindent
{\bf Proposition 4.} \emph{Suppose that assumptions {\bf A1'}-{\bf A3}, assumption {\bf B1}, assumption {\bf B2(ii)} for the first $p_{1}$ predictors, {\bf B2(ii)} for the next $(p_{2}-p_{1})$ predictors and {\bf B2(iii)} for the remaining $p-p_{2}$ predictors hold. {\bf (a)} Under the alternative hypothesis and $\textrm{M}$ fixed, we have} 
	\begin{align}
		{\mathcal S}_{\textrm{M}}(p_{0}) & \stackrel{d}\rightarrow \sum_{j=1}^{M} \left[
		{\cal Z}_{j}+
		\dfrac{\sqrt{4 p_{0}(1-p_{0})}}{|1-2p_{0}|} \frac{1}{\sigma_{\eta}} \		
		\left(\sum_{\ell=1}^{3} {\bm \delta}_{\ell}^{\top} {\bm Q}_{\ell,\infty} {\bm \delta}_{\ell}\right)
		\right]^{2}
		\label{eq:eq25}
	\end{align}
\emph{where $({\cal Z}_{1},\ldots,{\cal Z}_{M})$ denote \textrm{M} mutually independent standard normal random variables. {\bf (b)}  Under the alternative hypothesis $H_{1n}$, and $(n,M_{n})\rightarrow \infty$ such that $M_{n}/n \rightarrow 0$, we have}
\begin{align}
	{\cal Q}_{M_{n}}(p_{0}) & \stackrel{d}\rightarrow {\cal N}(0,1)+\frac{p_{0}(1-p_{0})}{(1-2p_{0})^{2} \ \sigma_{\eta}^{2}} \left(\sum_{\ell=1}^{3} \bm \delta^{\top}_{\ell} {\bm Q}_{\ell,\infty}\bm \delta_{\ell}\right)^{2}
	\label{eq:eq26}
\end{align}

\section{Tuning Parameters and Some Asymptotic Refinements}

The practical implementation of inferences based on ${\cal S}_{M}(p_{0})$ requires setting $p_{0}$ and $\textrm{M}$ in a way that ensures good size control while also delivering good power properties. As indicated by our findings in Proposition 2, for sufficiently large $n$, the specific choice of $p_{0}$ should minimally affect the size properties of the test statistics, provided that $p_{0}$ is bounded away from 0, 1, and 1/2, as required by assumption {\bf A1}. Consequently and in turn, our results in Propositions 3-4 point towards a strategy of setting $p_{0}$ at its tolerated boundary 
as $p_{0}=0.5-\epsilon_{b}$ for some small positive $\epsilon_{b}$ (e.g., $p_{0}=0.40$ or $p=0.45$). \\

It is important to reiterate the fact that 
our inferences remain valid for any $p_{0}$ in its permitted range. However, this flexibility also presents challenges in finite samples  where one may want to fine-tune the test statistic for maximum power without sacrificing size control.
Having previously argued in favor of setting $p_{0}$ at $p_{0}=1/2-\epsilon_{b}$, the objective of this section is to provide further theoretical justification for this strategy and to propose an appropriate magnitude for $\epsilon_{b}$.\\

Recalling that the domain of $p_{0}$ is given by 
${\cal P}_{\epsilon_{a},\epsilon_{b}}=\{p \colon 0<\epsilon_{a}\leq p\leq 1-\epsilon_{a}<1,|p-\frac{1}{2}|\geq \epsilon_{b}\}$ 
we first argue that setting $\epsilon_{a} \approx 0.3$, which implies $p_{0} \approx  [0.3,0.7] \setminus [0.5-\epsilon_{b},0.5+\epsilon_{b}]$, ensures excellent adequacy of the asymptotic approximations, resulting in remarkably stable and accurate size estimates. 
The intuition behind this point is that setting $p_{0}$ in this range ensures a low skewness of the distribution of $(w_{t}^{0}(p_{0})-1)\eta_{t}$, which asymptotically approaches zero by the Central Limit Theorem (CLT). \\

After establishing the insensitivity of the null distributions to this range of $p_{0}$, we proceed to investigate how close $p_{0}$ can be set to $1/2$ in order to maximize local power without inducing size distortions due to variance degeneracy at the $1/2$ boundary. We address this by examining the elasticity of the variance function with respect to $p$, which leads us to argue that setting $\epsilon_{b}=0.1$, or $p_{0}=0.4$, provides a good balance. 
Finally, we demonstrate that this choice of $p_{0}$ preserves the size properties of ${\cal S}_{M}(p_{0})$ regardless of alternative choices for $\textrm{M}$. \\

Before proceeding, however, we will address the important issue of why a data-driven approach is neither suitable not desirable in our context.

\subsection{Why not a data-driven approach?} 

In line with common practice in areas like long-run variance estimation and model specification, one may wish to argue in favour of a data-driven approach to setting $p_{0}$ and $\textrm{M}$. However, in our context such an approach is neither feasible nor advisable, as it is unlikely to produce valid inferences. The split-sample averaging we introduced within the classical Wald type formulation is essentially an artificial {\it device} intentionally designed to modify the asymptotics and make them amenable to detecting predictability in challenging environments. This mechanism deliberately distorts the classical asymptotics, and any data-driven method for setting $p_{0}$ would almost inevitably revert us back to the classical setting by steering $p_{0}$ toward its $1/2$ boundary. \\

To further understand this point, consider the analogy between our use of a split-sample average (instead of the standard sample mean) and the inclusion of threshold effects in regression models. Specifically, 
split-sample averaging can be likened to unnecessarily including threshold effects in a regression model - such as estimating $y_{t}=\beta_{1} I({\cal U}_{t}\leq p)+\beta_{2}I({\cal U}_{t}>p)+u_{t}$ instead of $y_{t}=\beta +u_{t}$ with ${\cal U}_{t}$ denoting a uniformly distributed i.i.d. sequence. Under appropriate assumption, estimating the threshold parameter would result in $\hat{p}$ converging in probability to $1/2$, i.e., the midpoint of the sample, and which is precisely the scenario we wish to rule out. \\

Some of the above points are also applicable to the IVX methodology which also requires a tuning parameter in its implementation and for which a data-driven approach for choosing its magnitude is not feasible. Instead, the authors 
recommend a suitable magnitude based on size/power simulations and which has proven to be highly effective. Any data-driven approach to determining the persistence parameter, which needs to be slightly less than one for generating instrumental variables, would likely point to one asymptotically.

\subsection{Influence of $p_{0}$ on the asymptotics}

We initially aim to better understand the robustness of the null asymptotics to $p_{0}$ by investigating whether 
there is a specific range of values of $p_{0}$ that are associated with a greater accuracy of the asymptotic approximations,  even if the underlying theory remains valid for any $p_{0} \in {\cal P}_{\epsilon_{a},\epsilon_{b}}$. 
Are there, for instance, magnitudes of 
$\epsilon_{a}$ that are more or less favourable? The first order asymptotics presented in in Propositions 1 and 2 do not address these concenrns directly. \\

To address these issues, we focus on the potential role of $p_{0}$
in the CLT used to establish the limiting distributions in Propositions 1 and 2. Since $p_{0}$ is our primary interest, we concentrate on ${\cal S}_{n}(p_{0})$ for $\textrm{M}=1$. From the proof of Proposition 1 we recall that the dominant term leading to the asymptotic normality of 
$\sqrt{n} \ \overline{d}_{n}$ was $\textrm{N}_{1n}(p_{0})$, reproduced below for convenience (see equation (\ref{eq:eq49}) in the appendix):
\begin{align}
	\textrm{N}_{1n}(p_{0}) & = \frac{1}{\sqrt{n}}\sum_{t=1}^{n}(w_{t}^{0}(p_{0})-1)\eta_{t}.
	\label{eq:eq27}
\end{align}
By invoking the CLT for martingale differences, we established that under the 
null hypothesis $\textrm{N}_{1n}(p_{0})\stackrel{d}\rightarrow N(0,V[w_{t}^{0}(p_{0})] E[\eta_{t}^{2}])$. However, this CLT-based result does not provide information on the rate at which $\textrm{N}_{1n}(p_{0})$ approaches normality. Such rates of convergence in CLTs are commonly referred to as Berry-Esseen bounds, which we exploit in a way that will help highlight the influence of $p_{0}$ on the accuracy of the CLT, as it applies to (\ref{eq:eq27}).  This allows us to evidence the fact that over a broad range of $p_{0}$ magnitudes such an influence should be negligible for sufficiently large $n$. \\

Since the early work of Berry-Esseen, extensive literature has focused on deriving such rates under different assumptions about the stochastic properties of processes involved. 
Berry-Esseen bounds provide a measure of the rate at which a statistic (typically a normalized sum of i.i.d. variables) converges to the standard normal distribution, offering an upper-bound on the difference between the CDF of a statistic and the standard normal CDF and which is particularly useful for understanding how quickly the CLT takes effect. In its classical form, under i.i.d. conditions, the Berry-Esseen bound for a suitably normalized and standardized sum of i.i.d. random variables $\{X_{i}\}$ is expressed as $C_{0} n^{-1/2} E|X_{1}|^{3}/(\sigma^{2}_{X})^{3/2}$, where $C_{0}$ is a constant independent of the sample size. This formulation highlights the significant role of the third absolute moment in determining the adequacy of the CLT-based normal approximation. While there is substantial literature on extending these bounds to dependent sequences, existing results often involve strong restrictions, such as the boundedness of martingale difference sequences, exponentially decaying mixing rates, and constraints on skewness (see El Machkouri and Ouchti (2007)). For example, in the context of strong mixing sequences, Lahiri and Sun (2009) discuss how a CLT convergence rate such as $n^{-1/2}$ may not be attainable.\\

In our context, although the terms in the summands of $\textrm{N}_{1n}(p_{0})$ are not i.i.d., we can leverage the independence of $(w_{t}^{0}(p_{0})-1)$ and $\eta_{t}$ to derive a suitable Berry-Esseen type bound, even if the $\eta_{t}'s$ are dependent (e.g., strong mixing or even long memory). Our result is summarized in the following proposition, where we defined $\sigma^{2}_{N} \coloneqq V[\textrm{N}_{1n}(p_{0})]=V[w_{t}(p_{0})]^{2}E[\eta_{t}^{2}]$ and $K_{0}$ is a universal constant that does not depend on $n$. \\

\noindent
{\bf Proposition 5.} \emph{Under assumptions {\bf A1} and {\bf A2} with $\kappa\geq 2$, we have 
	\begin{align}
		\sup_{x\in \mathds{R}}\left|P\left(\frac{\textrm{N}_{1n}(p_{0})}{\sigma_{N}}\leq x\right)-\Phi(x)\right| & \leq 
		\frac{K_{0}}{\sqrt{n}} \ g(p_{0}) \ \frac{1}{(\sigma^{2}_{\eta})^{3/2}} E|\eta_{1}|^{3},
		\label{eq:eq28}
	\end{align}
	where 
	\begin{align}
		g(p_{0}) & \coloneqq \frac{(1-2p_{0}(1-p_{0}))}{\sqrt{p_{0}(1-p_{0})}}
		\label{eq:eq29}
	\end{align}
}

Before interpreting this result, it is important to emphasize  that 
its validity relies on $p_{0} \in {\cal P}_{\epsilon_{a},\epsilon_{b}}$, as the asymptotics would naturally break down at the boundaries. It would therefore be misleading to view the result in (\ref{eq:eq28}) as specializing to a Berry-Esseen type bound for $\sum \eta_{t}/\sqrt{n}$ under $p_{0}=1/2$, even though $g(p_{0})=1$ at $p_{0}=1/2$. \\

The key observation here pertains to the shape of the function $g(p_{0})$ which is plotted in Figure 1. A close examination reveals its inverted-U shape with asymptotes that are parallel to the y-axis as $p_{0}$ approaches zero or one. 
\begin{figure}[h!]
	\centering
	\includegraphics[scale=0.8]{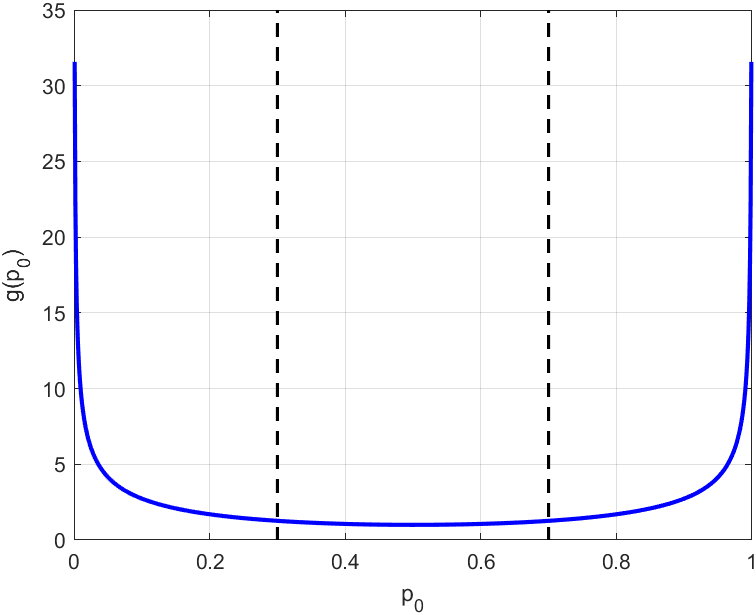}
	\caption{$g(p_{0})$}
	\label{fig:fig1}
\end{figure}
More importantly we can observe its low curvature for $p_{0} \approx [0.3,0.7]$ with $g(p_{0})$ changing only very mildly for such $p_{0}'s$. For example, for $p_{0} \in [0.3,0.7]$ we have $1.00<g(p_{0})\leq 1.3$ and for $p_{0} \in [0.35,0.65]$ we have $1.0<g(p_{0})<1.1$. 
These observations suggest that the asymptotic null distribution of our test statistic is likely to remain robust to a broad range of $p_{0}$
values, as long as $p_{0}$ is sufficiently far from the boundaries of zero and one. However, the case where $p_{0}=1/2$ is not fully captured by these arguments, as its impact operates via $\sigma_{N}$, which approaches degeneracy as $p_{0}$ nears $1/2$. Given the symmetry of the variance function, we expect minimal differences in the size properties of our tests for $p_{0}$ values in the range $\epsilon_{a} \approx 0.3$ to $1-\epsilon_{a} \approx 0.7$, subject to the constraint $|p-1/2|\geq \epsilon_{b}$. \\

\subsection{How close to 1/2 can we push $p_{0}$?}

Our earlier local power analysis pointed at setting $p_{0}$ at $p_{0}=1/2-\epsilon_{b}$ for an arbitrary small positive $\epsilon_{b}$. Bounding $p_{0}$ away from $1/2$ ensures a non-degenerate variance given by $(1-2p_{0})^{2}/(4p_{0}(1-p_{0})) \ \sigma^{2}_{\eta}$ for the limiting null distribution of $\textrm{N}_{1n}(p_{0})$. Consequently, and in finite samples, choosing an $\epsilon_{b}$ that is too small is likely to result in size distortions due to the test statistic concentrating too tightly around zero. The region of $p_{0}$ where this is most or least likely to occur can be gauged by examining the elasticity of $(1-2p_{0})^{2}/(4p_{0}(1-p_{0}))$ with respect to $p_{0}$, say $E(p_{0})$. This metric measures the percentage change in this asymptotic variance for a small percentage change in $p_{0}$ and is given by
\begin{align}
E(p_{0}) & =  \frac{-1}{(1-p_{0})(1-2p_{0})}.
	\label{eq:eq30}
\end{align}

A significant shift in elasticity around certain values of $p_{0}$ indicates the critical points where the variance becomes more or less sensitive to changes in $p_{0}$. 
\begin{figure}[h!]
	\centering
	\includegraphics[scale=0.8]{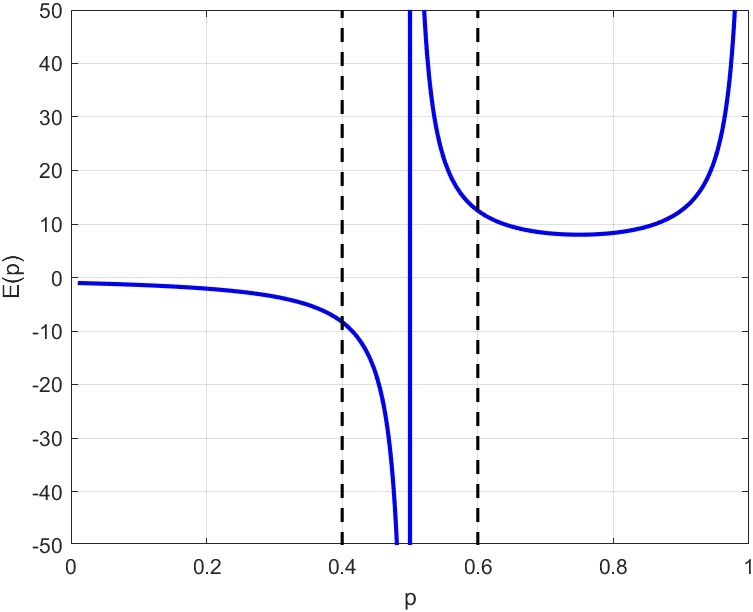}
	\caption{$E(p)$}
	\label{fig:fig2}
\end{figure}
From the plot of $E(p_{0})$ displayed in Figure 2 we can observe that $p_{0}$ in the vicinity of 0.4 (or 0.6 due to symmetry) is likely to offer a good compromise between the need to ensure good power and avoiding size distortions due to nearness to variance degeneracy. 

\subsection{Setting M: leveraging diminishing returns}

Our analysis thus far has not explicitly addressed the role of 
$\textrm{M}$, which becomes particularly significant in situations where predictors are purely stationary, and result in predictive regressions with low signal-to-noise ratios. As highlighted in Proposition 3, in such environments, and for a given significance level $\alpha$, asymptotic power is given by 
$P[\chi^{2}(\textrm{ncp}_{\textrm{M}};\textrm{M})>\chi^{2}(\textrm{M})_{\alpha}]$ where the noncentrality parameter $\textrm{ncp}_{\textrm{M}}=\textrm{M} \times \bm \lambda$ is as defined in equation (\ref{eq:eq24}). \\

A critical observation is that even under low signal to noise ratios as they occur in finance applications 
for instance, the magnitude of ${\bm \lambda}$ contributing to the noncentrality parameter often exceeds one, and frequently surpasses two or three. This in turn implies that beyond a typically low magnitude of $\textrm{M}$, marginal increases in power become negligible. The point is illustrated in Figure 3 below. \\

To illustrate these points further let us focus on the predictive regression  $y_{t}=\beta_{1}x_{1t-1}+u_{t}$ with a single predictor following the AR(1) process $x_{1t}=\phi_{1} x_{1t-1}+v_{1t}$, $|\phi|<1$. Straightforward algebra gives
\begin{align}
\textrm{ncp}_{M} & = \textrm{M} \ \underbrace{\frac{4p_{0}(1-p_{0})}{(1-2p_{0})^{2}} \ \delta_{1}^{4} \left(
\frac{\sigma^{2}_{v_{1}}}{\sigma^{2}_{u}}\right)^{2} \ \left(\frac{1}{(1-\phi_{1}^{2})^{2}}\right) \left(\frac{1}{{\sf Ku}(u_{t})-1}\right)}_{\bm \lambda}
	\label{eq:eq31}
\end{align}
where ${\sf Ku}(u_{t})$ denotes the kurtosis of $u_{t}$. It is now useful to link the above formulation with the $\textrm{SNR}=\beta_{1}^{2}E[x_{1t}^{2}]/\sigma^{2}_{u}=R^{2}/(1-R^{2})$ of the underlying predictive regression model, noting also that in this stationary predictor setting the local alternative is parameterized as $\delta_{1}=n^{1/4} \beta_{1}$. 
Setting $p_{0}=0.4$, $\phi_{1}=0.5$, $\sigma^{2}_{v_{1}}=1$ and ${\sf Ku}(u_{t})=3$ gives 
\begin{align}
{\bm \lambda} & = \frac{64}{3} \ 
\frac{\delta_{1}^{4}}{\sigma^{4}_{u}}.
	\label{eq:eq32}
\end{align}
illustrating the point that even under a very low $\textrm{R}^{2}$, we should expect $\bm \lambda\gg 1$. 
\begin{figure}[h!]
	\centering
	\includegraphics[scale=0.8]{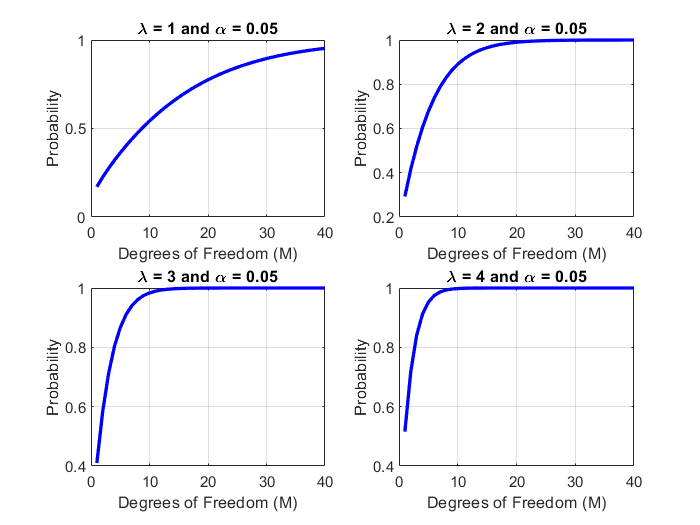}
	\caption{$P[\chi^{2}(M\lambda;M)>q_{\chi^{2}(M)}^{\alpha}]$}
	\label{fig:fig3}
\end{figure}

From the probabilities depicted in Figure 3 for instance we can infer that setting $\textrm{M}=5$ will typically be sufficient for predictive regressions with highly persistent predictors whereas in purely stationary contexts increasing $\textrm{M}$ up to 10-20 may still result in improved power. \\

When it comes to the test statistic ${\cal Q}_{M_{n}}(p_{0})$ the issue of attributing a value to $M$ does not arise explicitly since its limiting distribution does not depend on $M$ beyond requiring that $M_{n}/n\rightarrow 0$. This is akin to the issue of bandwidth specification in nonparameteric variance estimation where asymptotic theory only provides broad guidelines on how fast a bandwidth can be allowed to grow relative to the sample size $n$, and in practice bandwidths are often set as integer parts of fractional powers of $n$. This is also how we propose to implement our ${\cal Q}_{M_{n}}$ statistic when evaluating its finite sample properties below. Given the patterns illustrated in Figure 3 above and commonly encountered sample sizes in applied work, a choice such as $M_{n}=[n^{1/3}]$ is expected to offer a good size/power balance.

\section{Finite Sample Properties}

We investigate the finite sample size and power properties of our proposed approach to significance testing within a broad range of DGPs, starting with the most commonly encountered setting of a predictive regression with a single predictor ({\bf DGP1}). We subsequently consider more general environments with multiple predictors belonging to the same or different persistence classes ({\bf DGP2}). Existing methods for conducting inferences within {\it predictive regressions} have only rarely been put to test beyond single predictor settings. As we illustrate below favorable size/power properties in single predictor environments do not necessarily extend to multiple regression settings even if the latter involves only a few additional predictors. \\

In what follows we restrict our attention to the ${\cal Q}_{M_{n}}(p_{0})$ test statistic and present outcomes across a range of magnitudes for $p_{0}$ and specifying $M_{n}$ as $M_{n}=[n^{\delta}]$ for some $0<\delta<1$ such that $M_{n}/n\rightarrow 0$.

\subsection{DGP Parameterizations}
\noindent
{\bf DGP1}: we consider
\begin{align}
y_{t} & = \mu+\beta_{1}\ x_{1t-1}+u_{t} 
\label{eq:eq33}
\end{align}
where 
\begin{align}
x_{1t} & = \phi_{0}+\left(1-\dfrac{c_{1}}{n^{\alpha_{1}}}\right) x_{1t-1}+v_{1t}, 
\label{eq:eq34}
\end{align}
\begin{align}
u_{t} & = \rho \ u_{t-1}+\epsilon_{t},
\label{eq:eq35}
\end{align}
and
\begin{align}
	\epsilon_{t} & = \zeta_{t} \sqrt{\theta_{0}+\theta_{1} \epsilon_{t-1}^{2}}.
	\label{eq:eq36}
\end{align}
\noindent
Letting $w_{t}=(\zeta_{t},v_{1t})^{\top}$ the covariance matrix ${\bm \Omega}_{w}$ of the random disturbances is given by 
\begin{align}
{\bm \Omega}_{w} & =
\left(
\begin{array}{cc}
\sigma^{2}_{\zeta} & {\sigma}_{\zeta, v_{1}} \\
{\sigma}_{\zeta, v_{1}} & \sigma^{2}_{v_{1}}
\end{array}
\right)
	\label{eq:eq37}
\end{align}

\noindent and we impose $w_{t}\sim {\cal N}({\bm 0},{\bm \Omega}_{w})$ throughout all experiments. \\

We note that this setting can accommodate serial correlation, conditional heteroskedasticity as well as 
endogeneity with the latter operating through $\sigma_{\zeta, v_{1}}\neq 0$. The parameterizations that we consider aim to distinguish between three broad scenarios. (i) {\bf DGP1-(a)}: no serial correlation ($\rho=0$), conditional homoskedasticity ($\theta_{1}=0$) with possible endogeneity (i.e., $\sigma_{\zeta, v_{1}}\neq 0$ versus $\sigma_{\zeta, v_{1}}=0$). (ii) {\bf DGP1-(b)}: no serial correlation ($\rho=0$),  conditional heteroskedasticity ($\theta_{1}\neq 0$) with possible endogeneity (i.e., $\sigma_{\zeta, v_{1}}\neq 0$ versus $\sigma_{\zeta, v_{1}}=0$). (iii) {\bf DGP1-(c)}: serial correlation ($\rho \neq 0$),  conditional heteroskedasticity ($\theta_{1}\neq 0$) with possible endogeneity (i.e., $\sigma_{\zeta, v_{1}}\neq 0$ versus $\sigma_{\zeta, v_{1}}=0$).  \\

Specifically: 

\vspace{0.4cm}
\noindent
{\bf DGP1-(a)}: $(\rho,\theta_{0},\theta_{1},\sigma^{2}_{\zeta},\sigma^{2}_{v_{1}})=(0,2.5,0,1,1)$ with $\sigma_{\zeta, v_{1}} \in \{-0.90,0\}$, $\phi_{0}\in \{0,0.25\}$ 
\vspace{0.2cm}

\noindent
{\bf DGP1-(b)}: $(\rho,\theta_{0},\theta_{1},\sigma^{2}_{\zeta},\sigma^{2}_{v_{1}})=(0,2.5,0.25,1,1)$ with $\sigma_{\zeta, v_{1}} \in \{-0.90,0\}$,  $\phi_{0}\in \{0,0.25\}$ 
\vspace{0.2cm}

\noindent
{\bf DGP1-(c)}: $(\rho,\theta_{0},\theta_{1},\sigma^{2}_{\zeta},\sigma^{2}_{v_{1}})=(0.25,2.5,0.25,1,1)$ with $\sigma_{\zeta, v_{1}} \in \{-0.90,0\}$,  $\phi_{0}\in \{0,0.25\}$. \\

\noindent

These parameterizations imply an error variance of 
$\sigma_{u}^{2}=\theta_{0}=2.5$ under $\textrm{DGP1-(a)}$,  $\sigma_{u}^{2}=\theta_{0}/(1-\theta_{1})=3.33$ under  $\textrm{DGP1-(b)}$ and $\sigma^{2}_{u}=\theta_{0}/(1-\theta_{1})(1-\rho^{2})=3.56$ under $\textrm{DGP1-(c)}$. 
The combined effect of the ARCH(1) heteroskedasticity structure and the endogeneity-inducing non-zero $\sigma_{\zeta,v_{1}}$ yields a correlation $\rho_{u,v_{1}} \approx -0.90$ 
between the shocks to the predictor and predictand under $\rho=0$ (no serial correlation), and $\rho_{u,v_{1}}\approx -0.87$ under $\rho=0.5$ (serial correlation). Note that the ARCH(1) specification requires the restriction $3 \theta_{1}^{2}<1$ to ensure a finite fourth moment for $\epsilon_{t}$. Also, in the case of ${\bf DGP1-(c)}$, when $\alpha \approx 0$ (low to no persistence) is combined with serial-correlation ($\rho\neq 0$) and endogeneity, the least squares estimator is not consistent, violating our assumptions. We still include this scenario in the tables for completeness, and to help gauge how close to 1 the parameter $\alpha$ should be in order for consistency to kick in. \\

For these configurations, we examine a range of persistence characteristics by varying 
$\alpha_{1}\in \{0.00,0.25,0.50,0.75,0.95,1.00\}$. We maintain  $c_{1}=1$ throughout, except for the purely stationary case ($\alpha_{1}=0$), where we set $c_{1}=0.5$ to induce a stationary AR(1) process with a slope parameter of 0.5. In addition, we consider predictors with zero intercept $(\phi_{0}=0)$ 
and non-zero intercepts $(\phi_{0}\neq 0)$.  All estimated models include a fitted intercept, while the data-generating process (DGP) imposes $\mu=0$ with no loss of generality. \\

The size properties of test statistics are evaluated under $\beta_{1}=0$ in (\ref{eq:eq33}) using samples of size $n \in \{250,500, 1000\}$ across 10000 Monte-Carlo replications. Power is in turn evaluated under a fixed sample size of $n=500$, letting $\beta_{1}$ progressively depart from the null hypothesis. \\

\noindent
{\bf DGP2}: the second DGP aims to evaluate the sensitivity of inferences to the presence of multiple predictors and persistence heterogeneity. We consider  
\begin{align}
	y_{t} & = \mu+\beta_{1}\ x_{1t-1}+\beta_{2}\ x_{2t-1}+\beta_{3}\ x_{3t-1}+u_{t} 
	\label{eq:eq38}
\end{align}
where 
\begin{align}
	x_{it} & = \phi_{0,i} + \left(1-\dfrac{c_{i}}{n^{\alpha_{i}}}\right) x_{it-1}+v_{it} \ \ \ \ i=1,2,3,
	\label{eq:eq39}
\end{align}
\begin{align}
	u_{t} & = \rho u_{t-1}+\epsilon_{t}
	\label{eq:eq40}
\end{align}
with
\begin{align}
	\epsilon_{t} & = \zeta_{t} \sqrt{\theta_{0}+\theta_{1} \epsilon_{t-1}^{2}}.
	\label{eq:eq41}
\end{align}

\noindent
Letting $w_{t}=(\zeta_{t},v_{1t},v_{2t},v_{3t})^{\top}$ the $4\times 4$  covariance matrix ${\bm \Omega}_{w}$ of the random disturbances is given by 
\begin{align}
	{\bm \Omega}_{w} & =
	\left(
	\begin{array}{cc}
		\sigma^{2}_{\zeta} & {\bm \sigma}_{\zeta, v}^{\top} \\
		{\bm \sigma}_{\zeta, v} & {\bm \Sigma}_{vv}
	\end{array}
	\right)
	\label{eq:eq42}
\end{align}

\noindent and we impose $w_{t}\sim {\cal N}({\bm 0},{\bm \Omega}_{w})$ as before. 
We set
\begin{align}
	{\bm \Omega}_{w} & =
	\left(
	\begin{array}{cccc}
1.0350&	-0.9726 & -0.7408 &	-0.4943 \\
-0.9726	& 1.0214 &	0.5072	& 0.2545\\
-0.7408 &	0.5072&	1.0024 &	0.5015\\
-0.4943 & 0.2545 &	0.5015 &	1.0009
	\end{array}
	\right)
\label{eq:eq43}
\end{align}
and impose conditional heteroskedasticity across all DGPs, setting $(\theta_{0},\theta_{1})=(1.5,0.25)$ in (\ref{eq:eq41}). This latter parameterization results in $\sigma^{2}_{u}\approx 2.1$ under $\rho=0$ (no serial correlation) and $\sigma^{2}_{u}\approx 2.13$ under $\rho=0.25$ (serial correlation). 
$(\rho_{u,v_{1}},\rho_{u,v_{2}},\rho_{u,v_{3}}) \approx (-0.9,-0.7,-0.5)$ under $\rho=0$ (no serial correlation) \\


These DGPs are labelled as follows.\\

\noindent 
Stationary or Weakly Persistent Predictors:

\noindent 
{\bf DGP2-(a)}: $(\alpha_{1},\alpha_{2},\alpha_{3};\rho;\theta_{0},\theta_{1})=(0,0,0;0;1.5,0.25)$ \\
\noindent 
{\bf DGP2-(b)}: $(\alpha_{1},\alpha_{2},\alpha_{3};\rho;\theta_{0},\theta_{1})=(0.75,0.50,0.25;0;1.5,0.25)$ \\

\noindent 
Nearly Integrated Predictors: 

\noindent 
{\bf DGP2-(c)-(i)}: $(\alpha_{1},\alpha_{2},\alpha_{3};\rho;\theta_{0},\theta_{1})=(1,1,1;0;1.5,0.25)$ \\
\noindent 
{\bf DGP2-(c)-(ii)}: $(\alpha_{1},\alpha_{2},\alpha_{3};\rho;\theta_{0},\theta_{1})=(1,1,1;0.25;1.5,0.25)$ \\

\subsection{Empirical Size}

The empirical size outcomes for {\bf DGP1} ((a)-(c)) are presented in Tables \ref{tab:Table1}-\ref{tab:Table3}.  

\begin{itemize}
	
	\item[-] Across all DGP configurations, the robustness of ${\cal Q}_{M_{n}}(p_{0})$ with respect to varying magnitudes of $p_{0}$ is noteworthy, as predicted by the asymptotics. Nearly all entries of Tables \ref{tab:Table1}-\ref{tab:Table3} remain within 9\%-10\%, a fairly tight range around the nominal 10\%. Even for $p_{0}=0.42$, which might seem close to 0.5, the test's size remains near the nominal level across most scenarios. This suggests that mild proximity to 0.5 does not, in practice harm the test's finite sample behaviour. 
	This robustness of the proposed inferences holds regardless of endogeneity, serial correlation and/or conditional homoskedasticity. Our test statistic requires no corrections to address such distortions. Furthermore, the proposed test statistics 
	are robust to conditional heteroskedasticity across all values of $\alpha_{1}$, {\it including $\alpha_{1}=0$} (pure stationarity). While WIVX-based inferences also show robustness to conditional heteroskedasticity, this effect occurs solely when $\alpha$ is close to or equal to 1. \\
	
	\item[-] An especially important feature of ${\cal Q}_{M_{n}}$, evident in Tables \ref{tab:Table1}-\ref{tab:Table3}, is its robustness to the inclusion of a non-zero intercept in the predictor dynamics (i.e., $\phi_{0}\neq 0$). This is an issue that has received little attention in the literature on predictive regressions given the important distortions it causes for test statistics such as WIVX whose empirical sizes tend to zero in such instances. In Yang, Liu, Peng and Cai (2021), the authors conjecture that IVX based inferences fail to unify the cases of zero and nonzero intercepts and introduce a robust empirical likelihood based approach for conducting inferences in settings with intercepts in the covariate dynamics. \\


	\item[-] Under both serial correlation {\it and} conditional heteroskedasticity 
	 (Table \ref{tab:Table3}), the robustness of ${\cal Q}_{M_{n}}(p_{0})$ to $p_{0}$ is again evident.  When there is no endogeneity ($\rho_{u,v_{1}}=0$), the empirical sizes closely align with the nominal values across all $\alpha_{1}\in [0,1]$. However when serial correlation is combined with endogeneity (top panel of Table \ref{tab:Table3}), robustness is maintained only when predictors exhibit sufficient persistence, as illustrated in the top panel of Table \ref{tab:Table3} for $\alpha_{1}\geq 0.75$. This is in agreement with our discussion surrounding Assumptions {\bf B} as consistent estimation of model parameters is not feasible under endogeneity coupled with serial correlation.

\end{itemize}

\vspace{0.28cm}

The empirical size outcomes for {\bf DGP2} are presented in Table
\ref{tab:Table4} where we have also considered two alternative parameterizations for $M_{n}$ that satisfy $M_{n}/n\rightarrow 0$. The purpose of this second exercise is to assess test performance in contexts with multiple predictors. A common feature characterizing all test statistics is that in finite samples size properties deteriorate as we move from a single predictor to a multiple predictor setting as in DGP2.

\begin{itemize}
	
	\item[-] Focusing first on the top two panels of Table \ref{tab:Table4} which correspond to 
	DGPs with weakly persistent predictors ($(\alpha_{1},\alpha_{2},\alpha_{3})=(0,0,0)$ and $(\alpha_{1},\alpha_{2},\alpha_{3})=(0.75,0.50,0.25)$) we continue to note a good match between empirical and nominal sizes for the ${\cal Q}_{M_{n}}(p_{0})$ statistic. Under $p_{0}=0.4$ and for $n\geq 500$ for instance, 	all empirical sizes fall within the $[8\%,9.7\%]$ range. The corresponding range for WIVX is $[11.6\%,13.3\%]$.
	
	\item[-] When all predictors are nearly integrated ($(\alpha_{1},\alpha_{2},\alpha_{3})=(1,1,1)$) size distortions clearly kick in for smaller sample sizes but these can also be seen to approach the nominal size of 10\% as $n$ grows.
	Under $p_{0}=0.4$ the empirical sizes corresponding to ${\cal Q}_{M_{n}}(p_{0})$ lie in the range $[11.6\%,14.1\%]$ for $n>500$. 
	
	\item[-] The top 4 panels of Table \ref{tab:Table4} have implemented inferences using $M_{n}$ growing proportionately to $n^{1/3}$ while its bottom 4 panels repeat the same exercise by allowing $M_{n}$ to grow faster at rate $n^{1/2}$. Overall, from these results it is clear that an $M_{n}$ that grows too fast tends to result in oversized tests in smaller samples. 
	
	\item[-] Overall, these multiple-predictor based results highlight the fact that when the entire set of predictors are highly persistent our tests are moderately oversized in small to moderate sample sizes for magnitudes of $p_{0}$ near $0.4$ or above. For $p_{0} \approx 0.3$ however and under all scenarios involving $(\alpha_{1},\alpha_{2},\alpha_{3})=(1.00,1.00,1.00)$ empirical sizes closely match 10\% even for small sample sizes.

\end{itemize}

\subsection{Empirical Power}

We next investigate the power properties of ${\cal Q}_{M_{n}}(p_{0})$. We fix the sample size at $n=500$ and evaluate the frequencies of rejection of the null hypotheses as the slope parameters depart from the global null. For these experiments we restrict our attention to DGPs with no serial correlation and no heteroskedasticity. 
\begin{center}
	\begin{figure}[H]
		\centering
		\includegraphics[scale=0.68]{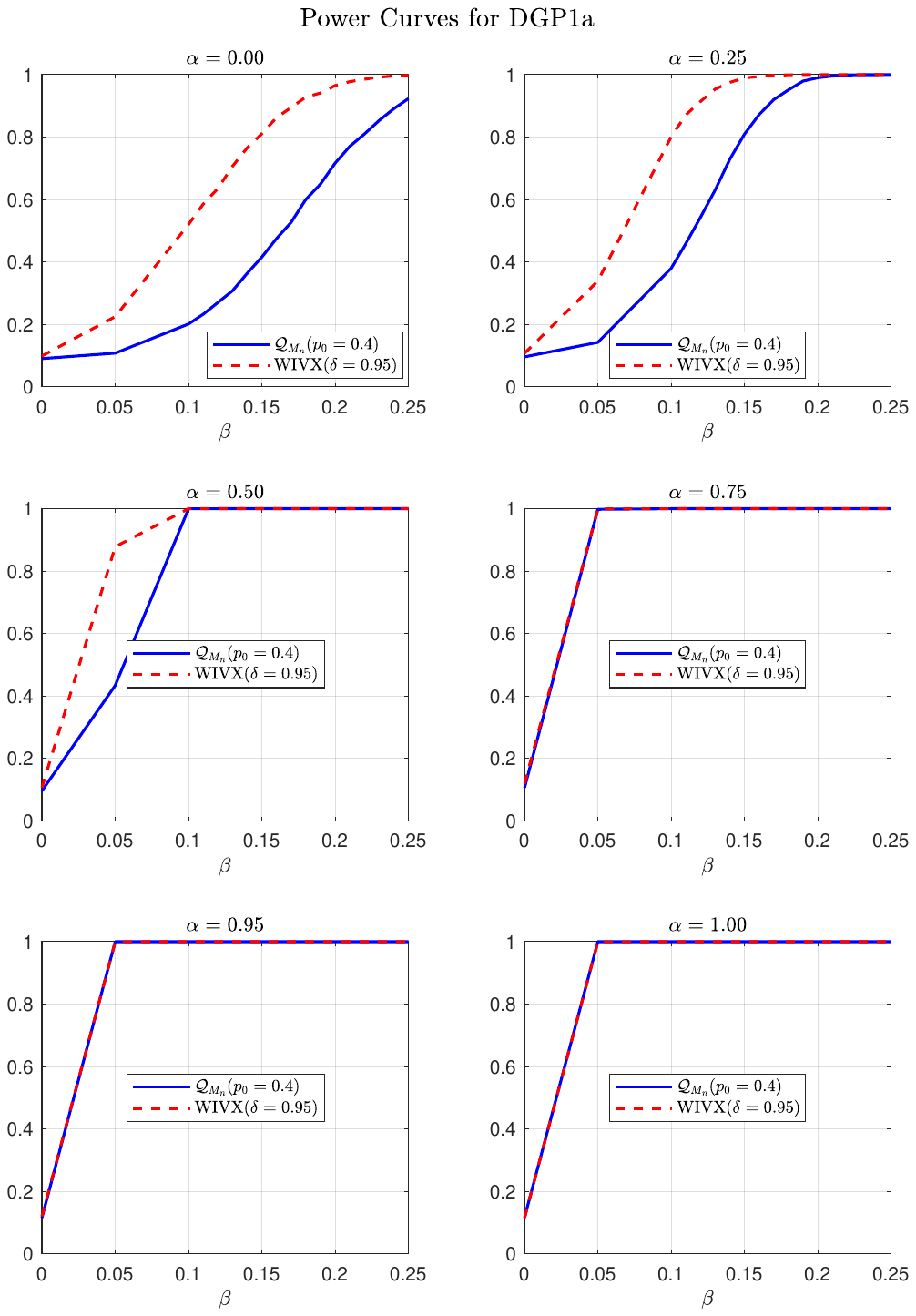}
		\caption{}
		\label{fig:fig4}
	\end{figure}
\end{center}

\FloatBarrier

\begin{center}
	\begin{figure}[H]
		\centering
		\includegraphics[scale=0.58]{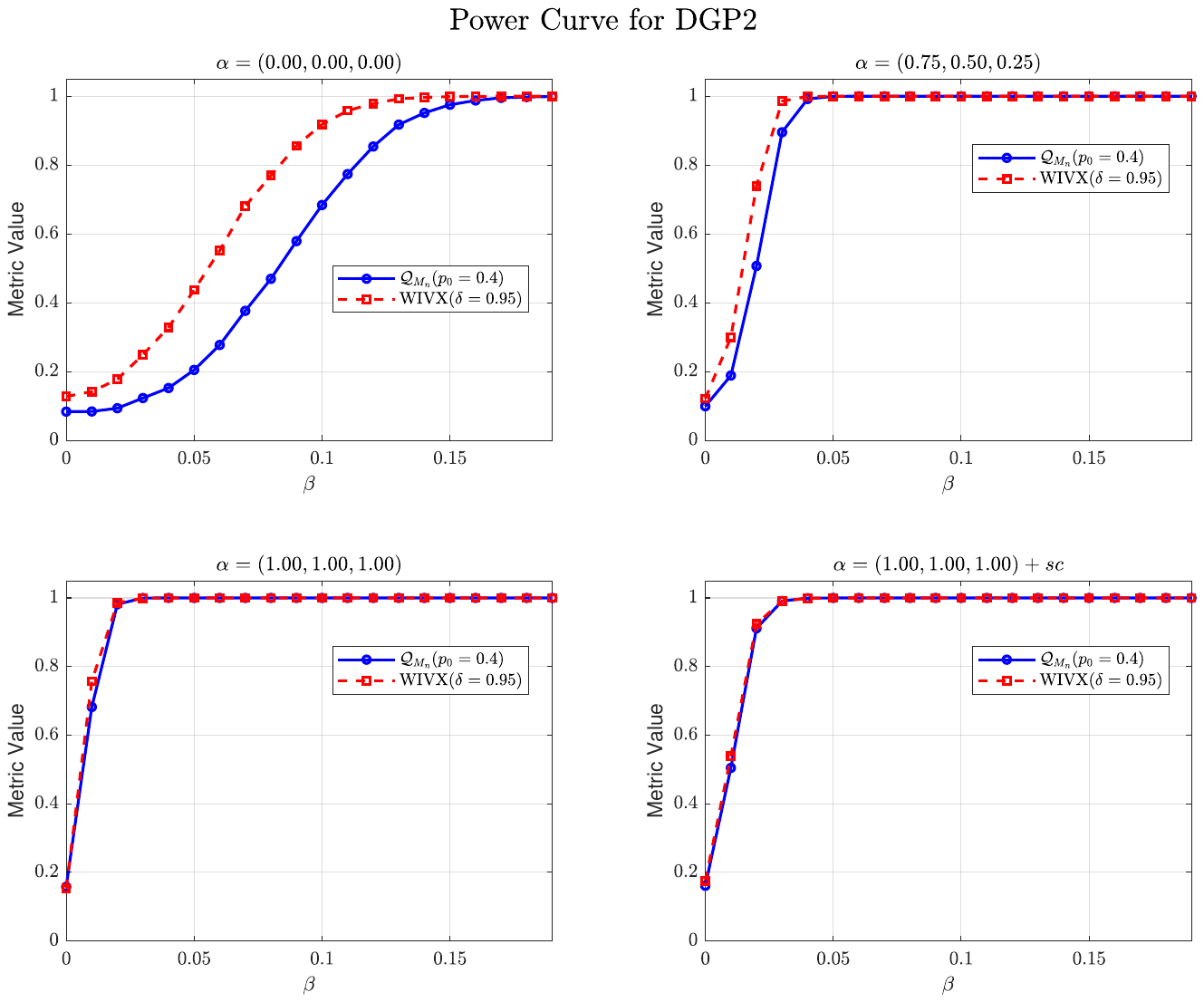}
		\caption{}
		\label{fig:fig5}
	\end{figure}
\end{center}

\FloatBarrier

For DGP1 corresponding to a single predictor scenario we vary $\beta$ from 0 (size) to 0.25 in increments of 0.01 with the resulting power profiles across alternative persistence characteristics,  presented in Figure 4 across different degrees of predictor persistence and for $p_{0}=0.4$. In instances where the predictor is purely stationary or has very low degrees of persistence (e.g., $\alpha \leq 0.5$) we note that inferences based on ${\cal Q}_{M_{n}}(p_{0})$ are dominated by WIVX based inferences. For $\alpha>0.5$ however we also note that both tests have nearly identical power profiles. 

Similar patterns also characterize the power profiles associated with $\textrm{DGP2}$ presented in Figure 5. Under all configurations whereby $\alpha_{i}\neq 0$ (i.e., mild to high persistence) ${\cal Q}_{M_{n}}(p_{0})$ is seen to share similar power properties with $\textrm{WIVX}$.

 \section{Conclusions}
 
 We have introduced a new approach for assessing the statistical significance of predictors in settings where serial dependence and different degrees of persistence  complicate standard predictive-regression inferences. By leveraging a split-sample device built around Bernoulli weights, the proposed methods achieve robustness to serial dependence and persistence without requiring special instruments, augmented regressions, or complex corrections for serial correlation and heteroskedasticity. The result is a straightforward Wald type test whose limit distributions are nuisance-parameter-free and remain valid under endogeneity, multiple predictors, and heterogeneous persistence. In finite samples our tests are shown to display excellent size properties under particularly challenging settings. However, our simulations have also shown that the cost of such a high degree of generality is a loss in power when predictors have very little to no persistence.

 \clearpage
 	\begin{center}
		{\bf Appendix: Proofs}
	\end{center}

\noindent
PROOF OF PROPOSITION 1. We initially establish the result in (\ref{eq:eq18}) for $\textrm{M}=1$. Given the regression model in (\ref{eq:eq2}) standard least squares algebra gives
$\hat{u}_{1t}^{2}=u_{t}^{2}+(\hat{\bm \theta}-{\bm \theta})^{\top}\widetilde{x}_{t-1}\widetilde{x}_{t-1}^{\top}(\hat{\bm \theta}-{\bm \theta})-2(\hat{\bm \theta}-{\bm \theta})^{\top}\widetilde{x}_{t-1}u_{t}$ 
for $\hat{\bm \theta}=(\sum \widetilde{\bm x}_{t-1} \widetilde{\bm x}_{t-1}')^{-1}
\sum \widetilde{\bm x}_{t-1} y_{t}$. Summing and applying suitable normalizations we obtain
\begin{align}
\dfrac{\sum \hat{u}_{1t}^{2}}{\sqrt{n}} & = \dfrac{\sum u_{t}^{2}}{\sqrt{n}}+\frac{1}{\sqrt{n}} [{\bm K}_{n}(\hat{\bm \theta}-{\bm \theta})]^{\top} {\bm K}_{n}^{-1}\sum_{t} \widetilde{x}_{t-1}\widetilde{x}_{t-1}^{\top} {\bm K}_{n}^{-1} [{\bm K}_{n}(\hat{\bm \theta}-{\bm \theta})] \nonumber \\
& - \frac{2}{\sqrt{n}} [{\bm K}_{n}(\hat{\bm \theta}-{\bm \theta})]^{\top} {\bm K}_{n}^{-1}\sum_{t} \widetilde{x}_{t-1}u_{t}.
\label{eq:eq44}
\end{align}
Assumptions ({\bf B1})-({\bf B2}) now ensure that 
\begin{align}
\dfrac{\sum \hat{u}_{1t}^{2}}{\sqrt{n}} & = \dfrac{\sum u_{t}^{2}}{\sqrt{n}}+o_{p}(1).
\label{eq:eq45}
\end{align}
Proceeding similarly for the restricted squared residuals under the null hypothesis we also obtain 
\begin{align}
	\dfrac{\sum \hat{u}_{0t}^{2}}{\sqrt{n}} & = \dfrac{\sum u_{t}^{2}}{\sqrt{n}}+o_{p}(1).
\label{eq:eq46}
\end{align}
Using (\ref{eq:eq45}) and (\ref{eq:eq46}) within (\ref{eq:eq7}) together with the fact that 
$\sum_{t=1}^{n}w_{t}(p_{0})=n$ we can express the numerator of ${\cal S}_{n}(p_{0})$ as
\begin{align}
\sqrt{n} \ \widetilde{N}_{n}(p_{0}) & =  \frac{1}{\sqrt{n}}\sum_{t} (w_{t}(p_{0})-1)\eta_{t}+o_{p}(1)
\label{eq:eq47}
\end{align}
\noindent where $\eta_{t}=u_{t}^{2}-E[u_{t}^{2}]$. 

\noindent
Using (\ref{eq:eq10})-(\ref{eq:eq11}) we have
\begin{align}
(w_{t}(p_{0})-1) & = (w_{t}^{0}(p_{0})-1) + \frac{b_{t}}{2}\left(\frac{1}{\overline{b}_{n}}-\frac{1}{p_{0}}\right)
+\frac{1-b_{t}}{2} \left(\frac{1}{1-\overline{b}_{n}}-\frac{1}{1-p_{0}}\right)
\label{eq:eq48}
\end{align} 
so that (\ref{eq:eq47}) can also be reformulated as 
\begin{align}
	\sqrt{n} \ \widetilde{N}_{n}(p_{0}) & =  \frac{1}{\sqrt{n}}\sum_{t} (w_{t}^{0}(p_{0})-1)\eta_{t}+
	\frac{1}{2}\left(\frac{1}{\overline{b}_{n}}-\frac{1}{p_{0}}\right)\frac{\sum_{t}b_{t}\eta_{t}}{\sqrt{n}} \nonumber \\
	& + \frac{1}{2}\left(\frac{1}{1-\overline{b}_{n}}-\frac{1}{1-p_{0}}\right)\frac{\sum_{t}(1-b_{t})\eta_{t}}{\sqrt{n}}+o_{p}(1) \nonumber \\
	& = \frac{1}{\sqrt{n}}\sum_{t} (w_{t}^{0}(p_{0})-1)\eta_{t}+\frac{1}{2}  \underbrace{\left(\frac{(1-2\overline{b}_{n})}{\overline{b}_{n}(1-\overline{b}_{n})}-\frac{(1-2p_{0})}{p_{0}(1-p_{0})}\right)}_{g_{1n}} \frac{\sum b_{t}\eta_{t}}{\sqrt{n}} \nonumber \\
	& + \frac{1}{2}\underbrace{\left(\frac{1}{1-\overline{b}_{n}}-\frac{1}{1-p_{0}}\right)}_{g_{2n}}\frac{\sum_{t}\eta_{t}}{\sqrt{n}}  \nonumber \\
& \coloneqq N_{1n}+N_{2n}+N_{3n}.
\label{eq:eq49}
\end{align}
\noindent
Given the i.i.d. nature of the $\{b_{t}\}$ sequence the strong law of large numbers ensures $\overline{b}_{n}\stackrel{as}\rightarrow p_{0}$ ($0<p_{0}<1$) and 
by the continuous mapping theorem it follows that $g_{1n}\stackrel{as}\rightarrow 0$ and $g_{2n}\stackrel{as}\rightarrow 0$. 
From assumption ({\bf A2}) we have $n^{-\frac{1}{2}} \sum_{t=1}^{n}\eta_{t}=O_{p}(1)$ and since $|b_{t}|\leq 1$ it also immediately follows that $n^{-\frac{1}{2}}\sum_{t=1}^{n}b_{t}\eta_{t}=O_{p}(1)$ so that $N_{2n}=o_{p}(1)$ and $N_{3n}=o_{p}(1)$ leading to 
\begin{align}
	\sqrt{n} \ \widetilde{N}_{n}(p_{0}) & =  \frac{1}{\sqrt{n}}\sum_{t} (w_{t}^{0}(p_{0})-1)\eta_{t}+o_{p}(1) \nonumber \\
	& \coloneqq \frac{1}{\sqrt{n}} \sum_{t=1}^{n} m_{t}(p_{0})+o_{p}(1).
\label{eq:eq50}
\end{align}

\noindent
To obtain the asymptotic distribution of (\ref{eq:eq50}) we next invoke the CLT for martingale difference sequences. To do this requires showing that $\{m_{t}(p_{0})\}$
is a martingale difference sequence with respect to a suitable filtration and that the associated Lindeberg condition holds. \\

\noindent
Let $\mathcal{F}_{t}=\sigma(\{b_{s},\eta_{s}\}_{s=1}^{t})$ and note that $E|m_{t}(p_{0})|=E|(w_{t}^{0}(p_{0})-1)\eta_{t}|\leq E|(w_{t}^{0}(p_{0})-1)|E|\eta_{t}|$. Since $(w_{t}^{0}(p_{0})-1)$ takes only two possible values it is bounded. Assumption ({\bf A2}) requiring 
$E|\eta_{t}|^{2+\delta}<\infty$ also ensures that $E|\eta_{t}|<\infty$. It therefore follows that  $E|m_{t}(p_{0})|<\infty$ as required for $m_{t}(p_{0})$ to be a candidate for a martingale difference sequence. Next, we have $E[m_{t}(p_{0})|\mathcal{F}_{t-1}]=E[(w_{t}^{0}(p_{0})-1)\eta_{t}|\mathcal{F}_{t-1}]=E[(w_{t}^{0}(p_{0})-1)|\mathcal{F}_{t-1}]E[\eta_{t}|\mathcal{F}_{t-1}]$. Clearly $E[(w_{t}^{0}(p_{0})-1)|\mathcal{F}_{t-1}]=0$. It thus remains to argue that under our assumptions the conditional expectation $E[\eta_{t}|\mathcal{F}_{t-1}]$ is finite (almost surely). Using Jensen's inequality for conditional expectations we have $|E[\eta_{t}|\mathcal{F}_{t-1}]|\leq E[|\eta_{t}||\mathcal{F}_{t-1}]$. Taking the expectation of both sides of the inequality and using the law of iterated expectations gives $E[|E[\eta_{t}|\mathcal{F}_{t-1}]|]\leq E[E[|\eta_{t}|\mathcal{F}_{t-1}]]=E|\eta_{t}|$. 
Since we operate under $E|\eta_{t}|^{2+\delta}<\infty$, $\eta_{t}$ is also integrable i.e., $E|\eta_{t}|<\infty$ which in turn implies that $E[|E[\eta_{t}|\mathcal{F}_{t-1}]|]$ must be finite and hence $E[\eta_{t}|\mathcal{F}_{t-1}]$ is also finite as required. We conclude that 
$m_{t}(p_{0})$ in (\ref{eq:eq50}) is a martingale difference sequence.  
The limiting distribution of $\sqrt{n}\widetilde{N}_{n}(p_{0})$ can now be established by invoking the CLT for martingale difference sequences provided that the convergence of the conditional variances of $m_{t}(p_{0})$ and the Lindeberg condition hold under our operating assumptions. \\

\noindent 
We initially consider the conditional variance of $m_{t}(p_{0})$. We note that $\eta_{t}^{2}$ is strictly stationary and ergodic due to $\eta_{t}$ being strictly stationary and ergodic. Jensen's inequality 
and the law of iterated expectations ensure that $E|E[\eta_{t}^{2}|\mathcal{F}_{t-1}]|\leq E[E[|\eta_{t}|^{2}|\mathcal{F}_{t-1}]]=E[\eta_{t}^{2}]<\infty$ since $E|\eta_{t}|^{2+\delta}<\infty$. It therefore follows from the mean ergodic theorem that
\begin{align}
\frac{1}{n}\sum_{t=1}^{n}E[\eta_{t}^{2}|\mathcal{F}_{t-1}] & \stackrel{p}\rightarrow \sigma^{2}_{\eta} \coloneqq E[\eta_{t}^{2}].
\label{eq:eq51}
\end{align}
\noindent
Recalling that $V[(w_{t}^{0}(p_{0})-1)]=(1-2p_{0})^{2}/4p_{0}(1-p_{0})$ and the independence of 
$(w_{t}^{0}(p_{0})-1)$ and $\eta_{t}$, we have 
\begin{align}
	\frac{1}{n}\sum_{t=1}^{n}E[m_{t}(p_{0})^{2}|\mathcal{F}_{t-1}] & = E[(w_{t}^{0}(p_{0})-1)^{2}] \frac{1}{n}\sum_{t=1}^{n}E[\eta_{t}^{2}|\mathcal{F}_{t-1}] \nonumber \\
	& = \frac{(1-2p_{0})^{2}}{4p_{0}(1-p_{0})} \frac{1}{n} \sum_{t=1}^{n}E[\eta_{t}^{2}|\mathcal{F}_{t-1}] \nonumber \\
	& \stackrel{p}\rightarrow \frac{(1-2p_{0})^{2}}{4p_{0}(1-p_{0})} \ E[\eta_{t}^{2}] \coloneqq v_{\infty}(p_{0}).
\label{eq:eq52}
\end{align}

\noindent
Next, the Lindeberg condition requires that for any $\epsilon>0$
\begin{align}
\frac{1}{n}\sum_{t=1}^{n} E[m_{t}(p_{0})^{2} {\mathds{1}}_{\{|m_{t}(p_{0})|>\epsilon \sqrt{n}\}}|\mathcal{F}_{t-1}] \stackrel{as}\rightarrow 0.
\label{eq:eq53}
\end{align}
\noindent
Given the i.i.d. nature of $(w_{t}^{0}(p_{0})-1)$ we can focus on verifying 
\begin{align}
\frac{1}{n}\sum_{t=1}^{n}(w_{t}^{0}(p_{0})-1)^{2}E[\eta_{t}^{2} \mathbbm{1}_{|m_{t}(p_{0})|>\epsilon \sqrt{n}}|\mathcal{F}_{t-1}] & \stackrel{as}\rightarrow 0.
\label{eq:eq54}
\end{align}
Appealing to the conditional H\"{o}lder inequality with conjugates $a=\frac{2+\kappa}{2}$ and $b=\frac{2+\kappa}{\kappa}$ gives 
\begin{align}
E[\eta_{t}^{2} \mathbbm{1}_{
\{|\eta_{t}|>\frac{\epsilon \sqrt{n}}{|w_{t}^{0}(p_{0})-1|}
\}
}|\mathcal{F}_{t-1}] &  \leq 
\left(
E[|\eta_{t}|^{2+\kappa}|\mathcal{F}_{t-1}]
\right)^{\frac{2}{2+\kappa}}
\left(
P[|\eta_{t}|>\frac{\epsilon \sqrt{n}}{|w_{t}^{0}(p_{0})-1|}|\mathcal{F}_{t-1}]
\right)^{\frac{\kappa}{2+\kappa}}.
\label{eq:eq55}
\end{align}
The second term in the right hand side of (\ref{eq:eq55}) can be bounded by appealing to the Markov inequality
\begin{align}
P\left(|\eta_{t}|>\frac{\epsilon \sqrt{n}}{|w_{t}^{0}(p_{0})-1|} \mid \mathcal{F}_{t-1}\right) & \leq 
\frac{E[|\eta_{t}|^{2+\kappa} \ \mid \mathcal{F}_{t-1}]}{
\left(\frac{\sqrt{n}\epsilon}{|w_{t}^{0}(p_{0})-1|}
\right)^{2+\kappa}
}.
\label{eq:eq56}
\end{align}
Plugging (\ref{eq:eq56}) back into (\ref{eq:eq55}) and noting that 
since $w_{t}^{0}(p_{0})$ is bounded there exists a constant $M>0$ such that $|w_{t}^{0}(p_{0})-1|\leq M$ it follows that
\begin{align}
E[\eta_{t}^{2} \mathbbm{1}_{
	\{|\eta_{t}|>\frac{\epsilon \sqrt{n}}{|w_{t}^{0}(p_{0})-1|}
	\}
} \ \mid \mathcal{F}_{t-1}] &  \leq 
\frac{M^{2+\kappa} E[|\eta_{t}|^{2+\kappa}|\mathcal{F}_{t-1}]}{n^{1+\frac{\kappa}{2}} \epsilon^{2+\kappa}}.
\label{eq:eq57}
\end{align}
\noindent Summing over $t$ and dividing by $n$ gives
\begin{align}
	\frac{1}{n}\sum_{t=1}^{n}(w_{t}^{0}(p_{0})-1)^{2}E[\eta_{t}^{2} \mathbbm{1}_{|m_{t}(p_{0})|>\epsilon \sqrt{n}}|\mathcal{F}_{t-1}] & \leq
	\frac{M^{2+\kappa}}{n^{\kappa/2} \epsilon^{2+\kappa}} \frac{1}{n} \sum_{t=1}^{n}E[|\eta_{t}|^{2+\kappa}|\mathcal{F}_{t-1}].
	\label{eq:eq58}
\end{align}
\noindent By the ergodic theorem and $E|\eta_{t}|^{2+\kappa}<\infty$ it follows that the right hand side of (\ref{eq:eq58}) tends to 0 almost surely. Since the left hand side is non negative we can therefore conclude that (\ref{eq:eq53}) holds, as required. The CLT for  martingale differences finally gives
 \begin{align}
 \sqrt{n} \widetilde{N}_{n}(p_{0}) & \stackrel{d}\rightarrow \mathcal{N}\left(0, \ \frac{(1-2p_{0})^{2}}{4p_{0}(1-p_{0})}E[\eta_{t}^{2}] \right).
 	\label{eq:eq59}
 \end{align}

\noindent
Next, letting $v_{\infty}(p_{0})\coloneqq ((1-2p_{0})/(4p_{0}(1-p_{0})))E[\eta_{t}^{2}]$
we write 
\begin{align}
\frac{\sqrt{n} \ \widetilde{N}_{n}(p_{0})}{s_{d}} & = 
\frac{\sqrt{n} \ \widetilde{N}_{n}(p_{0})}{\sqrt{v_{\infty}(p_{0})}} 
+
\frac{\sqrt{n} \ \widetilde{N}_{n}(p_{0})}{\sqrt{v_{\infty}(p_{0})}} 
\left(\frac{\sqrt{v_{\infty}(p_{0})}}{s_{d}}-1\right) \nonumber \\
& \coloneqq X_{1n} + X_{2n}.
 	\label{eq:eq60}
\end{align}

\noindent 
From (\ref{eq:eq59}) we have $X_{1n}\stackrel{d}\rightarrow {\cal N}(0,1)$ and $X_{1n}^{2}\stackrel{d}\rightarrow \chi^{2}_{1}$ so that we 
require that $X_{2n}\stackrel{p}\rightarrow 0$ holds. This will follow provided that $s_{d}^{2}\stackrel{p}\rightarrow v_{\infty}(p_{0})$ which we verify next. From (\ref{eq:eq45}) and (\ref{eq:eq46}) we have $\sum_{t=1}^{n} \hat{u}_{0t}^{2}/n=E[u_{t}^{2}]+o_{p}(1)$ and $\sum_{t=1}^{n} \hat{u}_{1t}^{2}/n=E[u_{t}^{2}]+o_{p}(1)$ so that $\sum_{t=1}^{n}d_{t}^{2}/n=n^{-1} \sum_{t=1}^{n} (w_{t}^{0}(p_{0})-1)^{2}\eta_{t}^{2}+o_{p}(1)$ and the mean ergodic theorem leads to 
$s_{d}^{2}\stackrel{p}\rightarrow v_{\infty}(p_{0})$. We therefore have that $X_{2n}\stackrel{p}\rightarrow 0$ and conclude that ${\cal S}_{n}(p_{0}) \stackrel{d}\rightarrow \chi^{2}_{1}$. \\

\noindent
Next, let ${\cal S}_{n,j}(p_{0})$ be constructed from $M$ independent Bernoulli-based processes $\{b_{j,t}\}$ $j=1,\ldots,M$, each driving its own martingale difference sequence $\{m_{j,t}(p_{0})\}_{t=1}^{n}$. Formally, for each fixed $j$, we have $E[m_{j,t}(p_{0})|{\cal F}_{j,t-1}]=0$, and the resulting {\it single-shot} statistic ${\cal S}_{n,j}(p_{0})$ is exactly the one-$j$ version analyzed above. From our previous arguments ${\cal S}_{n,j}(p_{0})\stackrel{d}\rightarrow \chi^{2}_{1}$ for each $j=1,\ldots,M$. The key to obtaining the joint convergence of $({\cal S}_{n,1}(p_{0}),\ldots,{\cal S}_{n,M}(p_{0}))$ is that all these sequences $\{m_{j,t}\}$ are mutually independent across $j$. Indeed, each $\{b_{j,t}\}$ is an i.i.d. Bernoulli sequence, and these Bernoulli sequences are independent across the different $j$. This ensures that 
${\cal S}_{n,1}(p_{0}),\ldots,{\cal S}_{n,M}(p_{0})$ have asymptotically independent limits. Formally, for any 
${\bm \lambda}=(\lambda_{1},\ldots,\lambda_{M})\in \mathds{R}^{M}$ define the linear combination
${\cal S}_{M}^{\lambda}(p_{0}) \coloneqq \sum_{j=1}^{M}\lambda_{j}{\cal S}_{n,j}(p_{0})$. Since ${\cal S}_{n,j}(p_{0})\stackrel{d}\rightarrow Z_{j}^{2}$, $Z_{j}\sim {\cal N}(0,1)$ and the $Z_{j}$ are independent by construction (due to the independence of the Bernoulli sequences and hence of the $\{m_{j,t}\}$), we get 
${\cal S}_{M}^{\lambda}(p_{0})\stackrel{d}\rightarrow \sum_{j=1}^{M}\lambda_{j}Z_{j}^{2}$ as $n\rightarrow \infty$. By the Cramer-Wold device, this one-dimensional convergence for every linear combination ${\bm \lambda}$ implies joint convergence $({\cal S}_{n,1}(p_{0}),\ldots,{\cal S}_{n,M}(p_{0}))\stackrel{d}\rightarrow (Z_{1}^{2},\ldots,Z_{M}^{2})$. In particular, summing over $j$ gives  ${\cal S}_{M}(p_{0})=\sum_{j=1}^{M}{\cal S}_{n,j}(p_{0})\stackrel{d}\rightarrow \sum_{j=1}^{M}Z_{j}^{2}\sim \chi^{2}_{M}$ as stated. \hfill $\blacksquare$. \\

\noindent
PROOF OF PROPOSITION 2. From the definitions of ${\cal S}_{M_{n}}(p_{0})$ and ${\cal Q}_{M_{n}}(p_{0})$ in (\ref{eq:eq16})-(\ref{eq:eq17}) we can express ${\cal Q}_{M_{n}}(p_{0})$ as a normalized sum of a triangular array 
\begin{align}
{\cal Q}_{M_{n}}(p_{0}) & = \frac{1}{\sqrt{M_{n}}} \sum_{j=1}^{M_{n}} \left(\frac{S_{n,j}(p_{0})-1}{\sqrt{2}}\right) \nonumber \\
& \coloneqq \frac{1}{\sqrt{M_{n}}} \sum_{j=1}^{M_{n}} X_{n,j}(p_{0}).
	\label{eq:eq61}
\end{align}
We recall from our results in Proposition 1 that $X_{n,j}(p_{0}) \stackrel{d}\rightarrow Y_{j} \sim (\chi^{2}_{1}-1)/\sqrt{2}$ for all $j=1,\ldots,M_{n}$ and where $Y_{j}$ has mean zero and variance 1. Given these properties we proceed showing that ${\cal Q}_{M_{n}}(p_{0})\stackrel{d}\rightarrow {\cal N}(0,1)$ by invoking a Lindeberg type CLT for triangular arrays. \\

\noindent First note that under the finite-moment assumptions in {\bf A1} (i.e., $\eta_{t}$ has finite $(2+\kappa)$-th moment, and the Bernoulli based weights are bounded), $\{{\cal S}_{n,j}(p_{0}),n=1,2,3,\ldots\}$ is uniformly integrable. Since ${\cal S}_{n,j}(p_{0})\stackrel{d}\rightarrow \chi^{2}_{1}$ (Proposition 1), we have $\lim_{n\rightarrow \infty}E[{\cal S}_{n,j}(p_{0})]=1$ 
and $\lim_{n\rightarrow \infty}V[{\cal S}_{n,j}(p_{0})]=2$. Hence it also holds that $E[X_{n,j}(p_{0})] \rightarrow 0$ and $V[X_{n,j}(p_{0})] \rightarrow 1$. These moment convergence 
hold for each fixed $j$ and as $n\rightarrow \infty$. However, due to the mutual independence of the $\{b_{j,t}\}$ and their i.i.d.'ness in $t$, all single-shot sequences ${\cal S}_{n,j}(p_{0})$ behave identically. Thus it also holds that $\sup_{1\leq j\leq M_{n}}|E[X_{n,j}(p_{0})]|\rightarrow 0$ and $\sup_{1\leq j\leq M_{n}}|V[X_{n,j}(p_{0})]-1|\rightarrow 0$. Hence, in large enough samples the sum $\sum_{j=1}^{M_{n}}X_{n,j}(p_{0})$ will be centered and have variance $M_{n}$. \\

\noindent
To invoke a suitable CLT for $\{X_{n,j}(p_{0})/\sqrt{M_{n}}: 1\leq j\leq M_{n}\}$ requires verifying that a Lindeberg type condition holds. Specifically, for any $\epsilon>0$, 
\begin{align}
\lim_{n\rightarrow \infty} \frac{1}{M_{n}}\sum_{j=1}^{M_{n}}E[X_{n,j}(p_{0})^{2}\mathbbm{1}_{\{|X_{n,j}(p_{0})|>\epsilon \sqrt{M_{n}}\}}] & = 0.
	\label{eq:eq62}
\end{align} 
must hold. For this,  it suffices to note  by appealing to the Markov inequality that for some $r>2$ we have
\begin{align}
\frac{1}{M_{n}} \sum_{j=1}^{M_{n}} E[X_{n,j}^{2}(p_{0})\mathbbm{1}_{\{|X_{n,j}(p_{0})|>\epsilon \sqrt{M_{n}}\}}] & \leq 
\frac{1}{M_{n}}\sum_{j=1}^{M_{n}} \frac{E|X_{n,j}(p_{0})|^{r}}{(\epsilon \sqrt{M_{n}})^{r-2}} \rightarrow 0
	\label{eq:eq63}
\end{align}
\noindent provided that $\sup_{n,j}E|X_{n,j}(p_{0})|^{r}<\infty$ which holds under {\bf A1}. Hence in a triangular array CLT setting $\sum_{j=1}^{M_{n}}X_{n,j}(p_{0})/\sqrt{M_{n}}$ will tend to a normal limit if the difference between this sum and a sum of i.i.d. (asymptotically) variables remains negligible. 
Specifically, defining 
$X_{n,j}(p_{0})=Y_{j}+r_{n,j}$ where $r_{n,j}\coloneqq X_{n,j}(p_{0})-Y_{j}$ we can write
\begin{align}
{\cal Q}_{M_{n}} & =  \frac{1}{\sqrt{M_{n}}}\sum_{j=1}^{M_{n}}Y_{j}+\frac{1}{\sqrt{M_{n}}} \sum_{j=1}^{M_{n}} r_{n,j}.
	\label{eq:eq64}
\end{align}
\noindent 
The classical CLT ensures that the first component in the right hand side of (\ref{eq:eq64}) converges in distribution to a standard normal random variable. Hence the stated result follows provided that $\sum_{j=1}^{M_{n}}r_{n,j}=o_{p}(\sqrt{M_{n}})$. Recall that these mutually independent $r_{n,j}'s$ are the remainders of the asymptotic $(\chi^{2}_{1}-1)/\sqrt{2}$ approximation of the $X_{n,j}(p_{0})$'s and are typically characterized by the uniform second-moment bound $\sup_{j}E|r_{n,j}|^{2}\leq C/\sqrt{n}$ so that the requirement whereby $M_{n}/n\rightarrow 0$ ensures that $\sum_{j=1}^{M_{n}}r_{n,j}/\sqrt{M_{n}}\stackrel{p}\rightarrow 0$ and therefore ${\cal Q}_{M_{n}}\stackrel{d}\rightarrow {\cal N}(0,1)$ as stated. \hfill $\blacksquare$. \\

\noindent
PROOF OF PROPOSITION 3. Due to the invariance of our test statistic to $\mu$ we set $\mu=0$ in the DGP. Under the alternative hypothesis it continues to be the case that 
$\sum_{t=1}^{n}\hat{u}_{1t}^{2}/\sqrt{n}=\sum_{t=1}^{n}u_{t}^{2}/\sqrt{n}+o_{p}(1)$ as in the proof of Proposition 1. The restricted residuals are now given by $\hat{u}_{0t}=u_{t}+{\bm \beta}_{n}^{\top}{\bm x}_{t-1}$ and $\hat{u}_{0t}^{2}=u_{t}^{2}+{\bm \beta}_{n}^{\top}{\bm x}_{t-1}{\bm x}_{t-1}^{\top}{\bm \beta}_{n}+2 {\bm \beta}_{n}^{\top}{\bm x}_{t-1}u_{t}$. Multiplying by the scalar sequence $w_{j,t}(p_{0})$, summing and normalizing throughout leads to 
\begin{align}
\frac{\sum_{t=1}^{n}w_{j,t}(p_{0})\hat{u}_{0t}^{2}}{\sqrt{n}} & = 
\frac{\sum_{t=1}^{n}w_{j,t}(p_{0})u_{t}^{2}}{\sqrt{n}}+
\frac{1}{\sqrt{n}} {\bm \beta}_{n}^{\top}\sum_{t=1}^{n} w_{j,t}(p_{0}) {\bm x}_{t-1} {\bm x}_{t-1}^{\top} {\bm \beta}_{n}+\frac{2}{\sqrt{n}} {\bm \beta}_{n}^{\top} \sum_{t=1}^{n}w_{j,t}(p_{0}) {\bm x}_{t-1}u_{t} \nonumber \\
& = \frac{\sum_{t=1}^{n}w_{j,t}(p_{0})u_{t}^{2}}{\sqrt{n}}+
{\bm \delta}^{\top} \overline{\bm K}_{n}^{-1} \sum_{t=1}^{n} w_{j,t}(p_{0}) {\bm x}_{t-1} {\bm x}_{t-1}^{\top} \overline{\bm K}_{n}^{-1} {\bm \delta} \nonumber \\
& +\frac{2}{n^{1/4}} \overline{\bm K}_{n}^{-1} {\bm \delta}^{\top} \sum_{t=1}^{n}w_{j,t}(p_{0}) {\bm x}_{t-1}u_{t}
	\label{eq:eq65}
\end{align}
Since $\hat{\sigma}^{2}\stackrel{p}\rightarrow E[u_{t}^{2}]$ we  write
\begin{align}
\frac{\sum_{t=1}^{n}w_{j,t}(p_{0})(\hat{u}_{0t}^{2}-\hat{\sigma}^{2})}{\sqrt{n}} & =  \frac{\sum_{t=1}^{n}w_{j,t}(p_{0})\eta_{t}}{\sqrt{n}}+
{\bm \delta}^{\top} \overline{\bm K}_{n}^{-1} \sum_{t=1}^{n} w_{j,t}(p_{0}) {\bm x}_{t-1} {\bm x}_{t-1}^{\top} \overline{\bm K}_{n}^{-1} {\bm \delta} \nonumber \\
& +\frac{2}{n^{1/4}} \overline{\bm K}_{n}^{-1} {\bm \delta}^{\top} \sum_{t=1}^{n}w_{j,t}(p_{0}) {\bm x}_{t-1}u_{t}+o_{p}(1) 	\label{eq:eq66}  \\
& = \frac{\sum_{t=1}^{n}w_{j,t}^{0}(p_{0})\eta_{t}}{\sqrt{n}}+
{\bm \delta}^{\top} \overline{\bm K}_{n}^{-1} \sum_{t=1}^{n} w_{j,t}^{0}(p_{0}) {\bm x}_{t-1} {\bm x}_{t-1}^{\top} \overline{\bm K}_{n}^{-1} {\bm \delta} \nonumber \\
& +\frac{2}{n^{1/4}} \overline{\bm K}_{n}^{-1} {\bm \delta}^{\top} \sum_{t=1}^{n}w_{j,t}^{0}(p_{0}) {\bm x}_{t-1}u_{t}+o_{p}(1) \label{eq:eq67} \\
& = \frac{\sum_{t=1}^{n}w_{j,t}^{0}(p_{0})\eta_{t}}{\sqrt{n}}+
{\bm \delta}^{\top} \overline{\bm K}_{n}^{-1} \sum_{t=1}^{n} {\bm x}_{t-1} {\bm x}_{t-1}^{\top} \overline{\bm K}_{n}^{-1} {\bm \delta} \nonumber \\
& +\frac{2}{n^{1/4}} \overline{\bm K}_{n}^{-1} {\bm \delta}^{\top} \sum_{t=1}^{n} {\bm x}_{t-1}u_{t}+o_{p}(1)  \label{eq:eq68} 
\end{align}
\noindent
where the last equality follows from the independence of the $w_{j,t}^{0}(p_{0})'s$ from the data and the fact that 
$E[w_{j,t}^{0}(p_{0})]=1$ $\forall j=1,\ldots,M$.  It now follows that 
$\sqrt{n}\widetilde{N}_{j,n}(p_{0})$ can be formulated as
\begin{align}
\sqrt{n}\ \widetilde{N}_{j,n}(p_{0}) 
& =\frac{\sum_{t=1}^{n}(w_{j,t}^{0}(p_{0})-1) \eta_{t}}{\sqrt{n}}+
{\bm \delta}^{\top} \overline{\bm K}_{n}^{-1} \sum_{t=1}^{n} {\bm x}_{t-1} {\bm x}_{t-1}^{\top} \overline{\bm K}_{n}^{-1} {\bm \delta} \nonumber \\
& +\frac{2}{n^{1/4}} \overline{\bm K}_{n}^{-1} {\bm \delta}^{\top} \sum_{t=1}^{n} {\bm x}_{t-1}u_{t}+o_{p}(1) \nonumber \\
& \coloneqq A_{1n}+A_{2n}+A_{3n}
	\label{eq:eq69}
\end{align}
 
From Proposition 1 we have $A_{1n}\stackrel{d}\rightarrow {\cal N}(0,v_{\infty}(p_{0}))$. Assumption ({\bf B1}) leads to $A_{3n}=O_{p}(n^{-\frac{1}{4}})$ and assumptions ({\bf B2(i)-(iii)}) give $A_{2n} \stackrel{p}\rightarrow {\bm \delta}^{\top} {\bm Q}_{\infty} {\bm \delta}$ for $Q_{\infty}$ given by either $Q_{1\infty}$, $Q_{2,\infty}$ or $Q_{3,\infty}$. Since $s_{j,d}^{2}(p_{0}) \stackrel{p}\rightarrow v_{\infty}(p_{0})$ it follows that 
\begin{align}
{\cal S}_{n,j}(p_{0}) & \stackrel{d}\rightarrow \left[
{\cal Z}_{j}+\frac{1}{\sqrt{v_{\infty}(p_{0j})}} {\bm \delta}^{\top} {\bm Q}_{\infty} {\bm \delta}\right]^{2}
	\label{eq:eq70}
\end{align}
so that the asymptotic independence of the ${\cal S}_{n,j}(p_{0})'s$ established in the proof of Proposition 1 leads to the result stated in (\ref{eq:eq21}). Part (b) of Proposition 3 follows very similar lines and is therefore omitted. \hfill $\blacksquare$ \\

\noindent
PROOF OF PROPOSITION 4. The stated results follow from identical steps to the ones established in the proof of Proposition 3. \hfill $\blacksquare$ \\

\noindent
PROOF OF PROPOSITION 5. We aim to show that $N_{1n}(p_{0})/\sigma_{N}$ where $\sigma_{N}=V[(w_{t}^{0}(p_{0})-1)\eta_{t}]$ satisfies a Berry-Esseen bound with rate $O(n^{-1/2})$. Let $\sigma^{2}_{w}=V[w_{t}^{0}(p_{0})-1]$,
$\sigma^{2}_{\eta}=V[\eta_{t}]$, $\sigma^{2}_{N}=\sigma^{2}_{w}\sigma^{2}_{\eta}$. By construction $(w_{t}^{0}(p_{0})-1)$ is i.i.d. and is also independent of the entire sequence $\{\eta_{t}\}$. Although $\eta_{t}$ is typically serially dependent and thus $(w_{t}^{0}(p_{0})-1)\eta_{t}$ is not i.i.d. for a direct evaluation of a Berry-Esseen bound for i.i.d. processes, we can leverage the i.i.d.'ness of $(w_{t}^{0}(p_{0})-1)$ via a suitable conditioning argument on the entire sequence $\{\eta_{t}\}$. \\

\noindent
Write $A_{t}\coloneqq (w_{t}^{0}(p_{0})-1)a_{t}$ 
for a fixed sample path $\{\eta_{t}=a_{t}\}$ so that (conditionally), $A_{t}$ is i.i.d., and $\sum_{t=1}^{n} A_{t}$ has mean 0 and variance 
$\sum_{t=1}^{n}a_{t}^{2}E[(w_{t}^{0}(p_{0})-1)^{2}]$. By a standard Berry-Esseen inequality for sums of i.i.d. mean zero random variables (see, e.g., Feller (1971)), there is a universal constant $K_{0}$ such that for all real $z$
\begin{align}
\sup_{z\in \mathds{R}}\left|
P\left(\frac{1}{\sqrt{n}}\sum_{t=1}^{n}A_{t}\leq z|\{\eta_{t}=a_{t}\}\right)-\Phi\left(z/\sqrt{V[\frac{1}{\sqrt{n}}\sum A_{t}]}\right)\right| & \leq 
\frac{K_{0}}{\sqrt{n}} \frac{E|A_{1}|^{3}}{(V[A_{1}])^{3/2}}
	\label{eq:eq71}
\end{align}
where $A_{1}=(w_{1}^{0}(p_{0})-1)a_{1}$. Since $(w_{1}^{0}(p_{0})-1)$ is independent of $a_{1}$ we get $E|A_{1}|^{3}=E|w_{1}^{0}(p_{0})-1|^{3}|a_{1}|^{3}$,
$V[A_{1}]=E[(w_{1}^{0}(p_{0})-1)^{2}] a_{1}^{2}=\sigma_{w}^{2}|a_{1}|^{2}$. Hence 
\begin{align}
\frac{E|A_{1}|^{3}}{(V[A_{1}])^{3/2}} & = 
\frac{E|w_{1}(p_{0})-1|^{3} |a_{1}|^{3}}{(\sigma^{2}_{w}|a_{1}|^{2})^{3/2}} \nonumber \\
& = \frac{E|w_{1}^{0}(p_{0})-1|^{3}}{(\sigma^{2}_{w})^{3/2}} |a_{1}|.
	\label{eq:eq72}
\end{align}
In turn $V[\sum_{t=1}^{n}A_{t}/\sqrt{n}]=\sigma^{2}_{w}\sum_{t=1}^{n}a_{t}^{2}/n$. Plugging these values into the Berry-Esseen bound and simplifying yields
\begin{align}
\sup_{z \in \mathds{R}}	\left|
	P\left(\frac{1}{\sqrt{n}}\sum_{t=1}^{n}A_{t} \leq z|\{a_{t}\}\right)-\Phi\left(z/\sqrt{\sigma^{2}_{w} \sum_{t=1}^{n}a_{t}^{2}/n}\right)\right| & \leq 
	\frac{K_{0}}{\sqrt{n}}
	\frac{E|w_{1}^{0}(p_{0})-1|^{3}}{(\sigma^{2}_{w})^{3/2}} \frac{\sum_{t=1}^{n}|a_{t}|^{3}}{(\sum_{t=1}^{n}a_{t}^{2})^{3/2}}.
		\label{eq:eq73}
			\end{align}
\noindent 
Integrating the above conditional bound with respect to the distribution of $\{\eta_{t}\}$ yields
\begin{align}
\sup_{z \in \mathds{R}} \left|
P(N_{1n}(p_{0})\leq z)-E\left[\Phi\left(\frac{z}{\sqrt{\sigma_{w}^{2}\frac{1}{n}
		\sum \eta_{t}^{2}}}\right)\right]\right| & \leq \frac{K_{0}}{\sqrt{n}} \frac{E|w_{t}^{0}(p_{0})-1|^{3}}{(\sigma^{2}_{w})^{3/2}} E\left[\frac{1}{(\sum_{t}\eta_{t}^{2})^{3/2}}\sum_{t=1}^{n}|\eta_{t}|^{3}
	\right].
		\label{eq:eq74}
\end{align}
\noindent
Since $\{\eta_{t}\}$ is strictly stationary and ergodic with finite $(2+\kappa)$-th moment ($\kappa>0$), the mean ergodic theorem implies $\sum_{t=1}^{n}\eta_{t}^{2}/n\stackrel{a.s}\rightarrow E[\eta_{1}^{2}]=\sigma^{2}_{\eta}$ and also 
$\sum_{t=1}^{n}|\eta_{t}|^{3}/n\stackrel{a.s}\rightarrow E|\eta_{1}|^{3}$. Since $\sigma^{2}_{N}=\sigma^{2}_{w}\sigma^{2}_{\eta}$ we arrive at 
\begin{align}
\sup_{z\in \mathds{R}} \left|
P(\frac{N_{1n}(p_{0})}{\sigma_{N}}\leq z)-\Phi(z)
\right| & \leq 
\frac{K_{0}}{\sqrt{n}} \frac{E|w_{t}^{0}(p_{0})-1|^{3}}{(\sigma^{2}_{w})^{3/2}} \frac{E|\eta_{1}|^{3}}{(\sigma^{2}_{\eta})^{3/2}} \nonumber \\
& = \frac{K_{0}}{\sqrt{n}} \ g(p_{0}) \ \frac{E|\eta_{1}|^{3}}{(\sigma^{2}_{\eta})^{3/2}}
	\label{eq:eq75}
\end{align}
as stated. \hfill $\blacksquare$ \\



\newpage
\begin{table}[htbp]
	\centering
	\caption{\footnotesize Empirical Size (10\% Nominal). {\bf DGP1(a)}: \small $y_{t}=\mu+\beta x_{t-1}+u_{t}$,   $x_{t}=\phi_{0}+(1-1/n^{\alpha})x_{t-1}+v_{t}$, $(u_{t},v_{t})^{\top} \sim \textrm{NID}(0,\Omega)$, $\Omega=\{\{2.5,\sigma_{u,v}\},\{\sigma_{u,v},1\}\}$, $\beta=0$, $\textrm{M}_{n}=\sqrt{n/p_{0}}$ (\footnotesize Homoskedasticity, No Serial Correlation).}
	\footnotesize 
	\setlength{\tabcolsep}{2.4pt} 
\scalebox{0.68}{\begin{tabular}{lcccccccccccccccccccc} \hline
		& \multicolumn{7}{c}{${\cal Q}_{M_{n}}(p_{0})$}         &       &       & \textrm{WIVX} & \multicolumn{7}{c}{${\cal Q}_{M_{n}}(p_{0})$}         &       &       & \textrm{WIVX} \\ \hline
		$p_{0}$ & 0.30  & 0.35  & 0.36  & 0.37  & 0.38  & 0.39  & 0.40  & 0.41  & 0.42  & $\delta=0.95$ & 0.30  & 0.35  & 0.36  & 0.37  & 0.38  & 0.39  & 0.40  & 0.41  & 0.42  & $\delta=0.95$ \\ \hline
		$\sigma_{u,v} = -0.90$ & \multicolumn{10}{c}{$\phi_{0}=0$}                                             & \multicolumn{10}{c}{$\phi_{0} \neq 0$} \\
		$\bm \alpha=\bf 0.00$ &       &       &       &       &       &       &       &       &       &       &       &       &       &       &       &       &       &       &       &  \\
		n = 250 & 0.095 & 0.091 & 0.085 & 0.087 & 0.087 & 0.088 & 0.093 & 0.095 & 0.102 & 0.103 & 0.107 & 0.089 & 0.091 & 0.086 & 0.090 & 0.087 & 0.091 & 0.095 & 0.101 & 0.096 \\
		n = 500 & 0.100 & 0.097 & 0.095 & 0.093 & 0.091 & 0.094 & 0.090 & 0.094 & 0.095 & 0.103 & 0.103 & 0.089 & 0.094 & 0.093 & 0.092 & 0.097 & 0.085 & 0.098 & 0.095 & 0.095 \\
		n = 1000 & 0.099 & 0.098 & 0.092 & 0.097 & 0.093 & 0.097 & 0.096 & 0.093 & 0.099 & 0.103 & 0.103 & 0.093 & 0.101 & 0.093 & 0.097 & 0.092 & 0.098 & 0.099 & 0.097 & 0.097 \\
		$\bm \alpha = \bf 0.25$ &       &       &       &       &       &       &       &       &       &       &       &       &       &       &       &       &       &       &       &  \\
		n = 250 & 0.108 & 0.088 & 0.089 & 0.086 & 0.090 & 0.091 & 0.088 & 0.094 & 0.105 & 0.108 & 0.096 & 0.089 & 0.088 & 0.086 & 0.084 & 0.088 & 0.096 & 0.098 & 0.105 & 0.095 \\
		n = 500 & 0.104 & 0.097 & 0.092 & 0.091 & 0.089 & 0.098 & 0.094 & 0.093 & 0.098 & 0.099 & 0.104 & 0.095 & 0.092 & 0.090 & 0.088 & 0.092 & 0.092 & 0.085 & 0.099 & 0.086 \\
		n = 1000 & 0.094 & 0.099 & 0.097 & 0.099 & 0.101 & 0.097 & 0.098 & 0.093 & 0.100 & 0.100 & 0.099 & 0.096 & 0.090 & 0.092 & 0.095 & 0.095 & 0.098 & 0.095 & 0.094 & 0.084 \\
		$\bm \alpha = \bf 0.50$ &       &       &       &       &       &       &       &       &       &       &       &       &       &       &       &       &       &       &       &  \\
		n = 250 & 0.101 & 0.094 & 0.086 & 0.093 & 0.097 & 0.087 & 0.098 & 0.101 & 0.105 & 0.114 & 0.099 & 0.090 & 0.090 & 0.089 & 0.090 & 0.094 & 0.094 & 0.104 & 0.104 & 0.052 \\
		n = 500 & 0.104 & 0.088 & 0.093 & 0.095 & 0.097 & 0.096 & 0.097 & 0.094 & 0.101 & 0.112 & 0.098 & 0.096 & 0.092 & 0.092 & 0.094 & 0.093 & 0.097 & 0.094 & 0.101 & 0.036 \\
		n = 1000 & 0.092 & 0.096 & 0.093 & 0.093 & 0.095 & 0.095 & 0.093 & 0.098 & 0.098 & 0.100 & 0.099 & 0.093 & 0.096 & 0.096 & 0.098 & 0.088 & 0.094 & 0.091 & 0.099 & 0.029 \\
		$\bm \alpha = \bf 0.75$ &       &       &       &       &       &       &       &       &       &       &       &       &       &       &       &       &       &       &       &  \\
		n = 250 & 0.100 & 0.094 & 0.096 & 0.097 & 0.103 & 0.095 & 0.107 & 0.113 & 0.118 & 0.123 & 0.101 & 0.094 & 0.092 & 0.097 & 0.095 & 0.099 & 0.104 & 0.110 & 0.109 & 0.007 \\
		n = 500 & 0.094 & 0.098 & 0.095 & 0.104 & 0.097 & 0.100 & 0.103 & 0.102 & 0.113 & 0.121 & 0.096 & 0.097 & 0.096 & 0.100 & 0.100 & 0.099 & 0.095 & 0.103 & 0.105 & 0.001 \\
		n = 1000 & 0.100 & 0.098 & 0.099 & 0.097 & 0.101 & 0.096 & 0.107 & 0.111 & 0.107 & 0.119 & 0.097 & 0.098 & 0.091 & 0.101 & 0.096 & 0.095 & 0.100 & 0.098 & 0.105 & 0.001 \\
		$\bm \alpha =\bf 0.95$ &       &       &       &       &       &       &       &       &       &       &       &       &       &       &       &       &       &       &       &  \\
		n = 250 & 0.096 & 0.100 & 0.100 & 0.100 & 0.105 & 0.110 & 0.115 & 0.123 & 0.130 & 0.131 & 0.103 & 0.098 & 0.087 & 0.092 & 0.100 & 0.094 & 0.095 & 0.105 & 0.108 & 0.002 \\
		n = 500 & 0.095 & 0.098 & 0.101 & 0.105 & 0.105 & 0.106 & 0.115 & 0.125 & 0.134 & 0.120 & 0.103 & 0.093 & 0.094 & 0.097 & 0.099 & 0.097 & 0.095 & 0.097 & 0.105 & 0.002 \\
		n = 1000 & 0.096 & 0.100 & 0.093 & 0.102 & 0.109 & 0.110 & 0.113 & 0.118 & 0.123 & 0.124 & 0.097 & 0.093 & 0.094 & 0.089 & 0.093 & 0.098 & 0.095 & 0.100 & 0.096 & 0.001 \\
		$\bm \alpha = \bf 1.00$ &       &       &       &       &       &       &       &       &       &       &       &       &       &       &       &       &       &       &       &  \\
		n = 250 & 0.097 & 0.099 & 0.101 & 0.108 & 0.108 & 0.110 & 0.122 & 0.125 & 0.138 & 0.120 & 0.102 & 0.093 & 0.092 & 0.095 & 0.099 & 0.094 & 0.101 & 0.099 & 0.109 & 0.003 \\
		n = 500 & 0.101 & 0.100 & 0.099 & 0.103 & 0.109 & 0.112 & 0.116 & 0.123 & 0.137 & 0.114 & 0.099 & 0.098 & 0.098 & 0.098 & 0.093 & 0.093 & 0.098 & 0.096 & 0.105 & 0.001 \\
		n = 1000 & 0.102 & 0.096 & 0.101 & 0.101 & 0.104 & 0.109 & 0.116 & 0.114 & 0.120 & 0.115 & 0.096 & 0.095 & 0.094 & 0.088 & 0.099 & 0.095 & 0.091 & 0.100 & 0.097 & 0.002 \\ \hline
		$\sigma_{u,v} = 0.0$ &       &       &       &       &       &       &       &       &       &       &       &       &       &       &       &       &       &       &       &  \\
		$\bm \alpha = \bf 0.00$ &       &       &       &       &       &       &       &       &       &       &       &       &       &       &       &       &       &       &       &  \\
		n = 250 & 0.099 & 0.090 & 0.088 & 0.090 & 0.087 & 0.088 & 0.085 & 0.094 & 0.106 & 0.102 & 0.102 & 0.094 & 0.089 & 0.091 & 0.090 & 0.091 & 0.092 & 0.096 & 0.100 & 0.098 \\
		n = 500 & 0.097 & 0.089 & 0.096 & 0.090 & 0.093 & 0.095 & 0.089 & 0.089 & 0.099 & 0.101 & 0.097 & 0.100 & 0.089 & 0.091 & 0.094 & 0.093 & 0.094 & 0.094 & 0.103 & 0.103 \\
		n = 1000 & 0.101 & 0.093 & 0.091 & 0.096 & 0.091 & 0.097 & 0.094 & 0.095 & 0.094 & 0.096 & 0.099 & 0.098 & 0.094 & 0.099 & 0.097 & 0.093 & 0.093 & 0.093 & 0.097 & 0.104 \\
		$\bm \alpha = \bf 0.25$ &       &       &       &       &       &       &       &       &       &       &       &       &       &       &       &       &       &       &       &  \\
		n = 250 & 0.101 & 0.090 & 0.087 & 0.094 & 0.090 & 0.088 & 0.091 & 0.101 & 0.099 & 0.099 & 0.104 & 0.088 & 0.090 & 0.090 & 0.091 & 0.084 & 0.092 & 0.090 & 0.101 & 0.106 \\
		n = 500 & 0.099 & 0.099 & 0.093 & 0.096 & 0.095 & 0.090 & 0.093 & 0.089 & 0.098 & 0.104 & 0.102 & 0.094 & 0.091 & 0.090 & 0.096 & 0.096 & 0.093 & 0.092 & 0.100 & 0.101 \\
		n = 1000 & 0.099 & 0.091 & 0.099 & 0.092 & 0.095 & 0.094 & 0.094 & 0.098 & 0.099 & 0.103 & 0.101 & 0.098 & 0.094 & 0.097 & 0.086 & 0.094 & 0.096 & 0.094 & 0.094 & 0.098 \\
		$\bm \alpha = \bf 0.50$ &       &       &       &       &       &       &       &       &       &       &       &       &       &       &       &       &       &       &       &  \\
		n = 250 & 0.101 & 0.094 & 0.089 & 0.086 & 0.089 & 0.087 & 0.088 & 0.100 & 0.106 & 0.105 & 0.098 & 0.093 & 0.088 & 0.086 & 0.088 & 0.089 & 0.092 & 0.096 & 0.098 & 0.102 \\
		n = 500 & 0.097 & 0.093 & 0.091 & 0.088 & 0.085 & 0.088 & 0.096 & 0.094 & 0.102 & 0.106 & 0.098 & 0.097 & 0.092 & 0.095 & 0.090 & 0.091 & 0.093 & 0.091 & 0.096 & 0.101 \\
		n = 1000 & 0.105 & 0.091 & 0.089 & 0.093 & 0.098 & 0.101 & 0.094 & 0.093 & 0.098 & 0.098 & 0.103 & 0.099 & 0.095 & 0.094 & 0.093 & 0.099 & 0.093 & 0.094 & 0.094 & 0.098 \\
		$\bm \alpha = \bf 0.75$ &       &       &       &       &       &       &       &       &       &       &       &       &       &       &       &       &       &       &       &  \\
		n = 250 & 0.103 & 0.089 & 0.090 & 0.094 & 0.083 & 0.090 & 0.087 & 0.103 & 0.098 & 0.101 & 0.104 & 0.083 & 0.092 & 0.089 & 0.092 & 0.089 & 0.091 & 0.090 & 0.098 & 0.090 \\
		n = 500 & 0.097 & 0.100 & 0.090 & 0.094 & 0.086 & 0.091 & 0.095 & 0.092 & 0.097 & 0.099 & 0.097 & 0.096 & 0.094 & 0.092 & 0.097 & 0.084 & 0.091 & 0.092 & 0.099 & 0.096 \\
		n = 1000 & 0.094 & 0.091 & 0.094 & 0.095 & 0.097 & 0.097 & 0.097 & 0.097 & 0.098 & 0.101 & 0.101 & 0.096 & 0.095 & 0.093 & 0.093 & 0.092 & 0.093 & 0.092 & 0.099 & 0.095 \\
		$ \bm \alpha = \bf 0.95$ &       &       &       &       &       &       &       &       &       &       &       &       &       &       &       &       &       &       &       &  \\
		n = 250 & 0.102 & 0.091 & 0.093 & 0.089 & 0.086 & 0.089 & 0.087 & 0.097 & 0.102 & 0.096 & 0.100 & 0.088 & 0.091 & 0.088 & 0.093 & 0.089 & 0.087 & 0.096 & 0.106 & 0.098 \\
		n = 500 & 0.104 & 0.093 & 0.097 & 0.092 & 0.096 & 0.095 & 0.091 & 0.089 & 0.100 & 0.102 & 0.102 & 0.088 & 0.086 & 0.093 & 0.088 & 0.093 & 0.094 & 0.090 & 0.101 & 0.095 \\
		n = 1000 & 0.094 & 0.093 & 0.095 & 0.093 & 0.098 & 0.095 & 0.099 & 0.097 & 0.097 & 0.099 & 0.102 & 0.094 & 0.097 & 0.096 & 0.094 & 0.095 & 0.096 & 0.094 & 0.096 & 0.094 \\
		$\bm \alpha =\bf  1.00$ &       &       &       &       &       &       &       &       &       &       &       &       &       &       &       &       &       &       &       &  \\
		n = 250 & 0.102 & 0.091 & 0.094 & 0.088 & 0.087 & 0.087 & 0.091 & 0.097 & 0.102 & 0.104 & 0.099 & 0.092 & 0.087 & 0.094 & 0.088 & 0.094 & 0.090 & 0.100 & 0.100 & 0.098 \\
		n = 500 & 0.097 & 0.094 & 0.091 & 0.095 & 0.094 & 0.087 & 0.096 & 0.088 & 0.100 & 0.097 & 0.095 & 0.095 & 0.093 & 0.091 & 0.091 & 0.090 & 0.094 & 0.090 & 0.098 & 0.097 \\
		n = 1000 & 0.102 & 0.099 & 0.093 & 0.094 & 0.093 & 0.096 & 0.091 & 0.093 & 0.098 & 0.103 & 0.096 & 0.097 & 0.092 & 0.094 & 0.092 & 0.097 & 0.098 & 0.097 & 0.098 & 0.093 \\
	\end{tabular}}
	\label{tab:Table1}
\end{table}

\begin{table}[htbp]
	\centering
	\caption{\footnotesize Empirical Size (10\% Nominal). {\bf DGP1(b)}: \small $y_{t}=\mu+\beta x_{t-1}+u_{t}$,   $x_{t}=\phi_{0}+(1-1/n^{\alpha})x_{t-1}+v_{t}$, $u_{t}=\zeta_{t} \sqrt{\theta_{0}+\theta_{1}u_{t-1}^{2}}$, $(\zeta_{t},v_{t})^{\top}\sim \textrm{NID}(0,\Omega)$, $\Omega=\{\{1,\sigma_{\zeta,v}\},\{\sigma_{\zeta,v},1\}\}$, $(\theta_{0},\theta_{1})=(2.5,0.25)$, $\beta=0$, $\textrm{M}_{n}=\sqrt{n/p_{0}}$ \footnotesize (Heteroskedasticity, No Serial Correlation)}
		\footnotesize 
	\setlength{\tabcolsep}{2.4pt} 
\scalebox{0.684}{\begin{tabular}{lcccccccccccccccccccc} \hline
		& \multicolumn{7}{c}{${\cal Q}_{M_{n}}(p_{0})$}         &       &       & \textrm{WIVX} & \multicolumn{7}{c}{${\cal Q}_{M_{n}}(p_{0})$}         &       &       & \textrm{WIVX} \\ \hline
		$p_{0}$ & 0.30  & 0.35  & 0.36  & 0.37  & 0.38  & 0.39  & 0.40  & 0.41  & 0.42  & $\delta=0.95$ & 0.30  & 0.35  & 0.36  & 0.37  & 0.38  & 0.39  & 0.40  & 0.41  & 0.42  & $\delta=0.95$ \\ \hline
		$\sigma_{\zeta,v_{1}} = -0.90$ & \multicolumn{10}{c}{$\phi_{0}=0$}                                             & \multicolumn{10}{c}{$\phi_{0} \neq 0$} \\
		$\bm \alpha = \bf 0.00$ &       &       &       &       &       &       &       &       &       &       &       &       &       &       &       &       &       &       &       &  \\
		n = 250 & 0.100 & 0.094 & 0.093 & 0.092 & 0.091 & 0.097 & 0.101 & 0.106 & 0.116 & 0.151 & 0.096 & 0.091 & 0.095 & 0.091 & 0.093 & 0.091 & 0.091 & 0.107 & 0.114 & 0.147 \\
		n = 500 & 0.096 & 0.089 & 0.088 & 0.090 & 0.088 & 0.092 & 0.093 & 0.099 & 0.105 & 0.157 & 0.100 & 0.094 & 0.091 & 0.092 & 0.089 & 0.093 & 0.097 & 0.100 & 0.105 & 0.141 \\
		n = 1000 & 0.101 & 0.093 & 0.094 & 0.102 & 0.091 & 0.092 & 0.093 & 0.098 & 0.099 & 0.147 & 0.106 & 0.089 & 0.091 & 0.094 & 0.094 & 0.092 & 0.098 & 0.096 & 0.099 & 0.148 \\
		$\bm \alpha = \bf 0.25$ &       &       &       &       &       &       &       &       &       &       &       &       &       &       &       &       &       &       &       &  \\
		n = 250 & 0.104 & 0.089 & 0.086 & 0.091 & 0.090 & 0.090 & 0.092 & 0.104 & 0.106 & 0.135 & 0.104 & 0.088 & 0.086 & 0.088 & 0.093 & 0.089 & 0.095 & 0.100 & 0.109 & 0.124 \\
		n = 500 & 0.104 & 0.088 & 0.095 & 0.092 & 0.089 & 0.091 & 0.085 & 0.092 & 0.099 & 0.133 & 0.107 & 0.087 & 0.089 & 0.090 & 0.093 & 0.090 & 0.089 & 0.091 & 0.100 & 0.122 \\
		n = 1000 & 0.099 & 0.092 & 0.092 & 0.094 & 0.092 & 0.094 & 0.092 & 0.090 & 0.096 & 0.126 & 0.102 & 0.094 & 0.094 & 0.092 & 0.096 & 0.093 & 0.089 & 0.099 & 0.097 & 0.109 \\
		$\bm \alpha = \bf 0.50$ &       &       &       &       &       &       &       &       &       &       &       &       &       &       &       &       &       &       &       &  \\
		n = 250 & 0.102 & 0.094 & 0.095 & 0.092 & 0.090 & 0.090 & 0.090 & 0.097 & 0.102 & 0.125 & 0.099 & 0.089 & 0.087 & 0.088 & 0.084 & 0.090 & 0.087 & 0.092 & 0.097 & 0.066 \\
		n = 500 & 0.099 & 0.096 & 0.094 & 0.094 & 0.088 & 0.084 & 0.090 & 0.091 & 0.094 & 0.113 & 0.102 & 0.085 & 0.092 & 0.093 & 0.098 & 0.091 & 0.092 & 0.091 & 0.095 & 0.048 \\
		n = 1000 & 0.106 & 0.093 & 0.095 & 0.091 & 0.089 & 0.091 & 0.098 & 0.089 & 0.092 & 0.109 & 0.106 & 0.092 & 0.089 & 0.091 & 0.098 & 0.094 & 0.093 & 0.092 & 0.094 & 0.033 \\
		$\bm \alpha = \bf 0.75$ &       &       &       &       &       &       &       &       &       &       &       &       &       &       &       &       &       &       &       &  \\
		n = 250 & 0.097 & 0.092 & 0.095 & 0.094 & 0.093 & 0.097 & 0.102 & 0.103 & 0.107 & 0.129 & 0.100 & 0.092 & 0.087 & 0.090 & 0.093 & 0.093 & 0.096 & 0.100 & 0.104 & 0.005 \\
		n = 500 & 0.103 & 0.096 & 0.099 & 0.099 & 0.095 & 0.095 & 0.097 & 0.098 & 0.106 & 0.120 & 0.105 & 0.091 & 0.091 & 0.093 & 0.092 & 0.094 & 0.094 & 0.094 & 0.102 & 0.002 \\
		n = 1000 & 0.104 & 0.093 & 0.092 & 0.095 & 0.095 & 0.097 & 0.097 & 0.098 & 0.102 & 0.122 & 0.109 & 0.092 & 0.099 & 0.095 & 0.093 & 0.091 & 0.094 & 0.091 & 0.095 & 0.001 \\
		$\bm \alpha =\bf 0.95$ &       &       &       &       &       &       &       &       &       &       &       &       &       &       &       &       &       &       &       &  \\
		n = 250 & 0.096 & 0.096 & 0.095 & 0.097 & 0.102 & 0.105 & 0.114 & 0.117 & 0.126 & 0.130 & 0.100 & 0.086 & 0.087 & 0.087 & 0.095 & 0.087 & 0.094 & 0.095 & 0.106 & 0.003 \\
		n = 500 & 0.099 & 0.092 & 0.093 & 0.099 & 0.090 & 0.103 & 0.109 & 0.109 & 0.119 & 0.119 & 0.099 & 0.090 & 0.093 & 0.092 & 0.090 & 0.090 & 0.096 & 0.095 & 0.104 & 0.001 \\
		n = 1000 & 0.101 & 0.101 & 0.095 & 0.101 & 0.095 & 0.097 & 0.101 & 0.110 & 0.113 & 0.114 & 0.103 & 0.095 & 0.092 & 0.096 & 0.091 & 0.089 & 0.097 & 0.093 & 0.097 & 0.001 \\
		$\bm \alpha = \bf 1.00$ &       &       &       &       &       &       &       &       &       &       &       &       &       &       &       &       &       &       &       &  \\
		n = 250 & 0.100 & 0.089 & 0.101 & 0.103 & 0.102 & 0.105 & 0.115 & 0.120 & 0.131 & 0.125 & 0.102 & 0.095 & 0.091 & 0.091 & 0.085 & 0.085 & 0.094 & 0.100 & 0.103 & 0.003 \\
		n = 500 & 0.099 & 0.094 & 0.095 & 0.094 & 0.097 & 0.106 & 0.111 & 0.113 & 0.120 & 0.124 & 0.108 & 0.095 & 0.095 & 0.090 & 0.091 & 0.089 & 0.093 & 0.090 & 0.098 & 0.001 \\
		n = 1000 & 0.096 & 0.096 & 0.094 & 0.104 & 0.104 & 0.099 & 0.104 & 0.105 & 0.111 & 0.119 & 0.099 & 0.098 & 0.095 & 0.093 & 0.095 & 0.094 & 0.092 & 0.094 & 0.096 & 0.001 \\ \hline
		$\sigma_{\zeta,v_{1}} = 0.0$ &       &       &       &       &       &       &       &       &       &       &       &       &       &       &       &       &       &       &       &  \\
		$\bm \alpha = \bf 0.00$ &       &       &       &       &       &       &       &       &       &       &       &       &       &       &       &       &       &       &       &  \\
		n = 250 & 0.110 & 0.083 & 0.090 & 0.084 & 0.085 & 0.086 & 0.085 & 0.092 & 0.099 & 0.104 & 0.105 & 0.087 & 0.085 & 0.084 & 0.081 & 0.080 & 0.081 & 0.089 & 0.095 & 0.097 \\
		n = 500 & 0.105 & 0.096 & 0.088 & 0.086 & 0.092 & 0.092 & 0.091 & 0.093 & 0.094 & 0.105 & 0.104 & 0.096 & 0.088 & 0.089 & 0.086 & 0.088 & 0.092 & 0.087 & 0.091 & 0.100 \\
		n = 1000 & 0.106 & 0.090 & 0.095 & 0.091 & 0.093 & 0.096 & 0.092 & 0.095 & 0.088 & 0.104 & 0.106 & 0.092 & 0.095 & 0.094 & 0.089 & 0.091 & 0.094 & 0.094 & 0.089 & 0.103 \\
		$\bm \alpha = \bf 0.25$ &       &       &       &       &       &       &       &       &       &       &       &       &       &       &       &       &       &       &       &  \\
		n = 250 & 0.104 & 0.086 & 0.088 & 0.085 & 0.087 & 0.085 & 0.088 & 0.084 & 0.095 & 0.107 & 0.104 & 0.089 & 0.087 & 0.085 & 0.088 & 0.083 & 0.089 & 0.088 & 0.099 & 0.100 \\
		n = 500 & 0.099 & 0.093 & 0.092 & 0.090 & 0.090 & 0.087 & 0.092 & 0.088 & 0.091 & 0.099 & 0.101 & 0.094 & 0.088 & 0.091 & 0.089 & 0.092 & 0.089 & 0.088 & 0.094 & 0.102 \\
		n = 1000 & 0.103 & 0.096 & 0.094 & 0.091 & 0.093 & 0.094 & 0.085 & 0.091 & 0.095 & 0.102 & 0.104 & 0.092 & 0.093 & 0.090 & 0.094 & 0.094 & 0.093 & 0.095 & 0.095 & 0.101 \\
		$\bm \alpha = \bf 0.50$ &       &       &       &       &       &       &       &       &       &       &       &       &       &       &       &       &       &       &       &  \\
		n = 250 & 0.102 & 0.097 & 0.087 & 0.093 & 0.092 & 0.088 & 0.086 & 0.096 & 0.095 & 0.101 & 0.100 & 0.088 & 0.087 & 0.086 & 0.086 & 0.083 & 0.085 & 0.090 & 0.097 & 0.100 \\
		n = 500 & 0.106 & 0.094 & 0.094 & 0.091 & 0.090 & 0.091 & 0.089 & 0.095 & 0.098 & 0.102 & 0.106 & 0.088 & 0.099 & 0.093 & 0.090 & 0.094 & 0.090 & 0.092 & 0.098 & 0.102 \\
		n = 1000 & 0.100 & 0.099 & 0.091 & 0.094 & 0.087 & 0.095 & 0.095 & 0.096 & 0.096 & 0.099 & 0.105 & 0.092 & 0.094 & 0.088 & 0.091 & 0.090 & 0.089 & 0.097 & 0.094 & 0.100 \\
		$\bm \alpha = \bf 0.75$ &       &       &       &       &       &       &       &       &       &       &       &       &       &       &       &       &       &       &       &  \\
		n = 250 & 0.102 & 0.091 & 0.084 & 0.085 & 0.089 & 0.084 & 0.087 & 0.091 & 0.094 & 0.102 & 0.108 & 0.093 & 0.087 & 0.087 & 0.084 & 0.087 & 0.087 & 0.085 & 0.098 & 0.097 \\
		n = 500 & 0.103 & 0.092 & 0.094 & 0.089 & 0.089 & 0.088 & 0.088 & 0.086 & 0.093 & 0.096 & 0.103 & 0.090 & 0.089 & 0.090 & 0.085 & 0.087 & 0.088 & 0.095 & 0.092 & 0.090 \\
		n = 1000 & 0.104 & 0.096 & 0.095 & 0.092 & 0.095 & 0.092 & 0.093 & 0.095 & 0.093 & 0.103 & 0.099 & 0.091 & 0.096 & 0.093 & 0.094 & 0.094 & 0.091 & 0.089 & 0.092 & 0.092 \\
		$\bm \alpha = \bf 0.95$ &       &       &       &       &       &       &       &       &       &       &       &       &       &       &       &       &       &       &       &  \\
		n = 250 & 0.097 & 0.090 & 0.093 & 0.084 & 0.085 & 0.084 & 0.087 & 0.092 & 0.096 & 0.104 & 0.104 & 0.090 & 0.091 & 0.089 & 0.089 & 0.083 & 0.083 & 0.087 & 0.097 & 0.096 \\
		n = 500 & 0.103 & 0.089 & 0.093 & 0.089 & 0.086 & 0.090 & 0.092 & 0.092 & 0.096 & 0.103 & 0.106 & 0.094 & 0.091 & 0.091 & 0.087 & 0.089 & 0.091 & 0.092 & 0.093 & 0.094 \\
		n = 1000 & 0.103 & 0.096 & 0.095 & 0.096 & 0.095 & 0.093 & 0.095 & 0.092 & 0.096 & 0.098 & 0.102 & 0.096 & 0.104 & 0.093 & 0.096 & 0.096 & 0.089 & 0.096 & 0.097 & 0.093 \\
		$\bm \alpha = \bf 1.00$ &       &       &       &       &       &       &       &       &       &       &       &       &       &       &       &       &       &       &       &  \\
		n = 250 & 0.104 & 0.090 & 0.088 & 0.083 & 0.089 & 0.088 & 0.088 & 0.095 & 0.096 & 0.108 & 0.102 & 0.091 & 0.091 & 0.091 & 0.088 & 0.081 & 0.090 & 0.093 & 0.093 & 0.094 \\
		n = 500 & 0.105 & 0.094 & 0.092 & 0.088 & 0.093 & 0.088 & 0.086 & 0.089 & 0.090 & 0.105 & 0.100 & 0.090 & 0.091 & 0.086 & 0.091 & 0.086 & 0.094 & 0.088 & 0.087 & 0.097 \\
		n = 1000 & 0.099 & 0.094 & 0.096 & 0.091 & 0.093 & 0.093 & 0.099 & 0.094 & 0.094 & 0.104 & 0.100 & 0.093 & 0.097 & 0.092 & 0.098 & 0.093 & 0.090 & 0.094 & 0.093 & 0.098 \\
	\end{tabular}}
	\label{tab:Table2}
\end{table}%

\begin{table}[htbp]
	\centering
	\caption{\footnotesize Empirical Size (10\% Nominal). {\bf DGP1(c)}: \small $y_{t}=\mu+\beta x_{t-1}+u_{t}$,  $u_{t}=\rho u_{t-1}+\epsilon_{t}$, $x_{t}=\phi_{0}+(1-1/n^{\alpha})x_{t-1}+v_{t}$, $\epsilon_{t}=\zeta_{t} \sqrt{\theta_{0}+\theta_{1}\epsilon_{t-1}^{2}}$, $(\zeta_{t},v_{t})^{\top}\sim \textrm{NID}(0,\Omega)$, $\Omega=\{\{1,\sigma_{\zeta,v}\},\{\sigma_{\zeta,v},1\}\}$, $(\theta_{0},\theta_{1})=(2.5,0.25)$, $\rho=0.25$, $\beta=0$, $\textrm{M}_{n}=\sqrt{n/p_{0}}$ \footnotesize (Heteroskedasticity, Serial Correlation)}
		\footnotesize 
	\setlength{\tabcolsep}{2.4pt} 
\scalebox{0.684}{\begin{tabular}{lcccccccccccccccccccc} \hline
		& \multicolumn{7}{c}{${\cal Q}_{M_{n}}(p_{0})$}         &       &       & \textrm{WIVX} & \multicolumn{7}{c}{${\cal Q}_{M_{n}}(p_{0})$}         &       &       & \textrm{WIVX} \\ \hline
		$p_{0}$ & 0.30  & 0.35  & 0.36  & 0.37  & 0.38  & 0.39  & 0.40  & 0.41  & 0.42  & $\delta=0.95$ & 0.30  & 0.35  & 0.36  & 0.37  & 0.38  & 0.39  & 0.40  & 0.41  & 0.42  & $\delta=0.95$ \\ \hline
		$\sigma_{\zeta,v_{1}} = -0.90$ & \multicolumn{10}{c}{$\phi_{0}=0$}                                             & \multicolumn{10}{c}{$\phi_{0} \neq 0$} \\
		$\bm \alpha = \bf 0.75$ &       &       &       &       &       &       &       &       &       &       &       &       &       &       &       &       &       &       &       &  \\
		n = 250 & 0.112 & 0.092 & 0.086 & 0.091 & 0.080 & 0.083 & 0.084 & 0.089 & 0.098 & 0.111 & 0.102 & 0.089 & 0.090 & 0.093 & 0.085 & 0.083 & 0.090 & 0.095 & 0.097 & 0.032 \\
		n = 500 & 0.110 & 0.095 & 0.096 & 0.089 & 0.092 & 0.088 & 0.088 & 0.094 & 0.095 & 0.129 & 0.102 & 0.093 & 0.093 & 0.093 & 0.091 & 0.093 & 0.097 & 0.094 & 0.103 & 0.025 \\
		n = 1000 & 0.106 & 0.099 & 0.099 & 0.093 & 0.090 & 0.094 & 0.095 & 0.097 & 0.094 & 0.154 & 0.101 & 0.093 & 0.099 & 0.096 & 0.104 & 0.100 & 0.101 & 0.098 & 0.098 & 0.018 \\
		$\bm \alpha = \bf 0.95$ &       &       &       &       &       &       &       &       &       &       &       &       &       &       &       &       &       &       &       &  \\
		n = 250 & 0.101 & 0.092 & 0.090 & 0.087 & 0.085 & 0.086 & 0.085 & 0.092 & 0.096 & 0.057 & 0.100 & 0.100 & 0.101 & 0.098 & 0.102 & 0.100 & 0.106 & 0.106 & 0.113 & 0.017 \\
		n = 500 & 0.103 & 0.091 & 0.092 & 0.091 & 0.088 & 0.086 & 0.090 & 0.100 & 0.089 & 0.051 & 0.104 & 0.099 & 0.095 & 0.100 & 0.104 & 0.101 & 0.109 & 0.109 & 0.114 & 0.010 \\
		n = 1000 & 0.111 & 0.098 & 0.096 & 0.090 & 0.091 & 0.090 & 0.089 & 0.090 & 0.093 & 0.052 & 0.105 & 0.102 & 0.102 & 0.098 & 0.102 & 0.101 & 0.105 & 0.105 & 0.107 & 0.007 \\
		$\bm \alpha = \bf 1.00$ &       &       &       &       &       &       &       &       &       &       &       &       &       &       &       &       &       &       &       &  \\
		n = 250 & 0.103 & 0.088 & 0.089 & 0.089 & 0.083 & 0.091 & 0.091 & 0.093 & 0.092 & 0.048 & 0.103 & 0.097 & 0.098 & 0.095 & 0.100 & 0.099 & 0.106 & 0.108 & 0.117 & 0.017 \\
		n = 500 & 0.095 & 0.094 & 0.093 & 0.091 & 0.090 & 0.085 & 0.087 & 0.089 & 0.094 & 0.040 & 0.100 & 0.097 & 0.098 & 0.095 & 0.100 & 0.100 & 0.102 & 0.107 & 0.113 & 0.011 \\
		n = 1000 & 0.101 & 0.097 & 0.089 & 0.091 & 0.091 & 0.094 & 0.097 & 0.094 & 0.096 & 0.042 & 0.102 & 0.089 & 0.099 & 0.100 & 0.100 & 0.104 & 0.104 & 0.114 & 0.118 & 0.008 \\ \hline
		$\sigma_{\zeta,v_{1}} = 0.0$ &       &       &       &       &       &       &       &       &       &       &       &       &       &       &       &       &       &       &       &  \\
		$\bm \alpha = \bf 0.00$ &       &       &       &       &       &       &       &       &       &       &       &       &       &       &       &       &       &       &       &  \\
		n = 250 & 0.103 & 0.096 & 0.090 & 0.089 & 0.092 & 0.090 & 0.088 & 0.097 & 0.100 & 0.140 & 0.104 & 0.086 & 0.091 & 0.089 & 0.094 & 0.095 & 0.088 & 0.102 & 0.105 & 0.150 \\
		n = 500 & 0.101 & 0.095 & 0.092 & 0.096 & 0.092 & 0.092 & 0.090 & 0.098 & 0.104 & 0.149 & 0.102 & 0.095 & 0.091 & 0.099 & 0.096 & 0.094 & 0.099 & 0.095 & 0.103 & 0.148 \\
		n = 1000 & 0.107 & 0.097 & 0.101 & 0.093 & 0.090 & 0.097 & 0.098 & 0.098 & 0.100 & 0.147 & 0.109 & 0.094 & 0.098 & 0.097 & 0.097 & 0.094 & 0.091 & 0.099 & 0.099 & 0.149 \\
		$\bm \alpha = \bf 0.25$ &       &       &       &       &       &       &       &       &       &       &       &       &       &       &       &       &       &       &       &  \\
		n = 250 & 0.099 & 0.092 & 0.098 & 0.096 & 0.101 & 0.090 & 0.101 & 0.106 & 0.112 & 0.186 & 0.106 & 0.094 & 0.091 & 0.093 & 0.100 & 0.097 & 0.096 & 0.105 & 0.111 & 0.178 \\
		n = 500 & 0.105 & 0.096 & 0.103 & 0.098 & 0.097 & 0.103 & 0.107 & 0.099 & 0.109 & 0.179 & 0.100 & 0.095 & 0.102 & 0.103 & 0.098 & 0.100 & 0.100 & 0.102 & 0.110 & 0.180 \\
		n = 1000 & 0.100 & 0.096 & 0.096 & 0.100 & 0.096 & 0.100 & 0.102 & 0.102 & 0.105 & 0.181 & 0.102 & 0.098 & 0.098 & 0.093 & 0.101 & 0.100 & 0.101 & 0.108 & 0.104 & 0.186 \\
		$\bm \alpha = \bf  0.50$ &       &       &       &       &       &       &       &       &       &       &       &       &       &       &       &       &       &       &       &  \\
		n = 250 & 0.103 & 0.099 & 0.097 & 0.100 & 0.103 & 0.103 & 0.105 & 0.104 & 0.112 & 0.186 & 0.106 & 0.099 & 0.102 & 0.102 & 0.105 & 0.105 & 0.102 & 0.110 & 0.118 & 0.201 \\
		n = 500 & 0.109 & 0.096 & 0.095 & 0.092 & 0.097 & 0.102 & 0.102 & 0.112 & 0.116 & 0.188 & 0.101 & 0.102 & 0.094 & 0.100 & 0.103 & 0.101 & 0.102 & 0.104 & 0.112 & 0.197 \\
		n = 1000 & 0.102 & 0.094 & 0.100 & 0.098 & 0.097 & 0.102 & 0.099 & 0.106 & 0.113 & 0.202 & 0.107 & 0.097 & 0.100 & 0.097 & 0.099 & 0.107 & 0.107 & 0.105 & 0.107 & 0.197 \\
		$\bm \alpha = \bf  0.75$ &       &       &       &       &       &       &       &       &       &       &       &       &       &       &       &       &       &       &       &  \\
		n = 250 & 0.105 & 0.094 & 0.100 & 0.096 & 0.104 & 0.103 & 0.109 & 0.114 & 0.117 & 0.195 & 0.107 & 0.097 & 0.102 & 0.100 & 0.101 & 0.110 & 0.101 & 0.111 & 0.122 & 0.193 \\
		n = 500 & 0.106 & 0.106 & 0.101 & 0.105 & 0.107 & 0.101 & 0.109 & 0.113 & 0.124 & 0.204 & 0.107 & 0.093 & 0.099 & 0.095 & 0.096 & 0.105 & 0.111 & 0.110 & 0.111 & 0.192 \\
		n = 1000 & 0.100 & 0.098 & 0.094 & 0.094 & 0.101 & 0.106 & 0.108 & 0.111 & 0.109 & 0.195 & 0.100 & 0.101 & 0.097 & 0.100 & 0.096 & 0.098 & 0.105 & 0.105 & 0.107 & 0.190 \\
		$\bm \alpha = \bf 0.95$ &       &       &       &       &       &       &       &       &       &       &       &       &       &       &       &       &       &       &       &  \\
		n = 250 & 0.100 & 0.097 & 0.097 & 0.099 & 0.099 & 0.103 & 0.110 & 0.110 & 0.125 & 0.196 & 0.107 & 0.097 & 0.105 & 0.099 & 0.108 & 0.105 & 0.107 & 0.113 & 0.118 & 0.194 \\
		n = 500 & 0.101 & 0.101 & 0.093 & 0.099 & 0.103 & 0.101 & 0.102 & 0.103 & 0.118 & 0.199 & 0.104 & 0.101 & 0.106 & 0.103 & 0.106 & 0.108 & 0.105 & 0.108 & 0.119 & 0.186 \\
		n = 1000 & 0.103 & 0.095 & 0.094 & 0.098 & 0.102 & 0.102 & 0.101 & 0.104 & 0.112 & 0.202 & 0.105 & 0.100 & 0.096 & 0.099 & 0.099 & 0.096 & 0.102 & 0.104 & 0.111 & 0.199 \\
		$\bm \alpha = \bf 1.00$ &       &       &       &       &       &       &       &       &       &       &       &       &       &       &       &       &       &       &       &  \\
		n = 250 & 0.102 & 0.102 & 0.098 & 0.100 & 0.101 & 0.099 & 0.108 & 0.111 & 0.120 & 0.198 & 0.107 & 0.099 & 0.099 & 0.096 & 0.102 & 0.101 & 0.102 & 0.110 & 0.114 & 0.187 \\
		n = 500 & 0.102 & 0.098 & 0.103 & 0.106 & 0.104 & 0.107 & 0.109 & 0.114 & 0.115 & 0.200 & 0.101 & 0.102 & 0.099 & 0.098 & 0.101 & 0.102 & 0.110 & 0.108 & 0.116 & 0.192 \\
		n = 1000 & 0.109 & 0.099 & 0.099 & 0.101 & 0.104 & 0.104 & 0.108 & 0.110 & 0.114 & 0.202 & 0.105 & 0.102 & 0.097 & 0.098 & 0.098 & 0.103 & 0.104 & 0.105 & 0.114 & 0.193 \\
	\end{tabular}}
	\label{tab:Table3}
\end{table}

\begin{table}[htbp]
	\centering
	\caption{Empirical Size (10\% Nominal)}
		\footnotesize 
	\setlength{\tabcolsep}{3pt} 
\scalebox{0.88}{\begin{tabular}{lcccccccccc} \hline
		$M_{n}=[(n/p_{0})^{1/3}]$ & \multicolumn{9}{c}{${\cal Q}_{M_{n}}(p_{0})$}                         & $\textrm{WIVX}$ \\ \hline
		& \multicolumn{10}{c}{$\textrm{DGP2-(a)}: (\bm \alpha_{1},\bm \alpha_{2},\bm \alpha_{3})=(\bf 0.00,\bf 0.00,\bf 0.00)$, $\textrm{Het+No Serial Correlation}$} \\ \hline
		$p_{0}$ & 0.30  & 0.35  & 0.36  & 0.37  & 0.38  & 0.39  & 0.40  & 0.41  & 0.42  & $\delta=0.95$ \\ \hline
		n = 250 & 0.068 & 0.070 & 0.071 & 0.069 & 0.069 & 0.072 & 0.074 & 0.070 & 0.073 & 0.138 \\
		n = 500 & 0.074 & 0.072 & 0.077 & 0.075 & 0.081 & 0.081 & 0.084 & 0.087 & 0.086 & 0.133 \\
		n = 1000 & 0.086 & 0.083 & 0.079 & 0.081 & 0.083 & 0.083 & 0.088 & 0.086 & 0.088 & 0.127 \\
		n = 2000 & 0.084 & 0.086 & 0.086 & 0.084 & 0.087 & 0.085 & 0.090 & 0.089 & 0.090 & 0.126 \\ \hline
		& \multicolumn{10}{c}{$\textrm{DGP2-(b)}: (\bm \alpha_{1},\bm  \alpha_{2},\bm \alpha_{3})=(\bf 0.75,\bf 0.50,\bf 0.25)$, $\textrm{Het+No Serial Correlation}$} \\
		n = 250 & 0.079 & 0.088 & 0.089 & 0.095 & 0.097 & 0.104 & 0.111 & 0.106 & 0.108 & 0.138 \\
		n = 500 & 0.078 & 0.079 & 0.081 & 0.084 & 0.090 & 0.095 & 0.095 & 0.099 & 0.106 & 0.124 \\
		n = 1000 & 0.083 & 0.084 & 0.082 & 0.088 & 0.086 & 0.089 & 0.092 & 0.100 & 0.102 & 0.113 \\
		n = 2000 & 0.083 & 0.085 & 0.081 & 0.084 & 0.086 & 0.086 & 0.097 & 0.093 & 0.099 & 0.116 \\ \hline
		& \multicolumn{10}{c}{$\textrm{DGP2-c-(i)}: (\bm \alpha_{1},\bm  \alpha_{2},\bm \alpha_{3})=(\bf 1.00,\bf 1.00,\bf 1.00)$, $\textrm{Het+No Serial Correlation}$} \\
		n = 250 & 0.095 & 0.122 & 0.130 & 0.137 & 0.153 & 0.163 & 0.178 & 0.178 & 0.189 & 0.168 \\
		n = 500 & 0.088 & 0.106 & 0.113 & 0.122 & 0.134 & 0.150 & 0.152 & 0.176 & 0.185 & 0.155 \\
		n = 1000 & 0.085 & 0.096 & 0.102 & 0.105 & 0.110 & 0.123 & 0.139 & 0.149 & 0.163 & 0.145 \\
		n = 2000 & 0.088 & 0.089 & 0.095 & 0.100 & 0.105 & 0.110 & 0.116 & 0.127 & 0.133 & 0.145 \\ \hline
		& \multicolumn{10}{c}{$\textrm{DGP2-c-(ii)}: (\bm \alpha_{1},\bm \alpha_{2},\bm \alpha_{3})=(\bf 1.00,\bf 1.00,\bf 1.00)$,  $\textrm{Het+Serial Correlation}$} \\
		n = 250 & 0.097 & 0.131 & 0.139 & 0.152 & 0.156 & 0.167 & 0.177 & 0.183 & 0.193 & 0.188 \\
		n = 500 & 0.089 & 0.116 & 0.118 & 0.127 & 0.140 & 0.150 & 0.157 & 0.176 & 0.185 & 0.180 \\
		n = 1000 & 0.086 & 0.104 & 0.101 & 0.113 & 0.119 & 0.126 & 0.141 & 0.153 & 0.162 & 0.173 \\
		n = 2000 & 0.085 & 0.096 & 0.100 & 0.104 & 0.112 & 0.118 & 0.130 & 0.139 & 0.152 & 0.182 \\ \hline\hline \hline
		$M_{n}=[(n/p_{0})^{1/2}]$ & \multicolumn{10}{c}{$\textrm{DGP2-(a)}: (\bm \alpha_{1},\bm \alpha_{2},\bm \alpha_{3})=(\bf 0.00,\bf 0.00,\bf 0.00)$, $\textrm{Het+No Serial Correlation}$} \\
		n = 250 & 0.096 & 0.106 & 0.108 & 0.114 & 0.120 & 0.128 & 0.140 & 0.154 & 0.161 & 0.135 \\
		n = 500 & 0.099 & 0.100 & 0.103 & 0.105 & 0.116 & 0.118 & 0.125 & 0.136 & 0.143 & 0.130 \\
		n = 1000 & 0.104 & 0.102 & 0.100 & 0.097 & 0.108 & 0.110 & 0.122 & 0.120 & 0.132 & 0.128 \\
		n = 2000 & 0.092 & 0.099 & 0.098 & 0.098 & 0.097 & 0.101 & 0.112 & 0.116 & 0.120 & 0.131 \\ \hline
		& \multicolumn{10}{c}{$\textrm{DGP2-(b)}: (\bm \alpha_{1},\bm \alpha_{2},\bm \alpha_{3})=(\bf 0.75,\bf 0.50,\bf 0.25)$, $\textrm{Het+No Serial Correlation}$} \\
		n = 250 & 0.105 & 0.127 & 0.133 & 0.140 & 0.147 & 0.157 & 0.168 & 0.180 & 0.194 & 0.130 \\
		n = 500 & 0.094 & 0.109 & 0.113 & 0.124 & 0.132 & 0.141 & 0.145 & 0.166 & 0.176 & 0.122 \\
		n = 1000 & 0.099 & 0.102 & 0.104 & 0.107 & 0.115 & 0.126 & 0.134 & 0.140 & 0.166 & 0.115 \\
		n = 2000 & 0.098 & 0.103 & 0.100 & 0.103 & 0.104 & 0.109 & 0.118 & 0.120 & 0.134 & 0.115 \\ \hline
		& \multicolumn{10}{c}{$\textrm{DGP2-c-(i)}: (\bm \alpha_{1},\bm \alpha_{2},\bm \alpha_{3})=(\bf 1.00,\bf 1.00,\bf 1.00)$, $\textrm{Het+No Serial Correlation}$} \\
		n = 250 & 0.129 & 0.189 & 0.201 & 0.212 & 0.238 & 0.251 & 0.269 & 0.283 & 0.297 & 0.163 \\
		n = 500 & 0.119 & 0.157 & 0.182 & 0.184 & 0.203 & 0.227 & 0.243 & 0.256 & 0.285 & 0.155 \\
		n = 1000 & 0.109 & 0.127 & 0.141 & 0.154 & 0.169 & 0.192 & 0.206 & 0.228 & 0.251 & 0.148 \\
		n = 2000 & 0.097 & 0.124 & 0.123 & 0.130 & 0.139 & 0.158 & 0.165 & 0.194 & 0.212 & 0.146 \\ \hline
		& \multicolumn{10}{c}{$\textrm{DGP2-c-(ii)}: (\bm \alpha_{1},\bm \alpha_{2},\bm \alpha_{3})=(\bf 1.00,\bf 1.00,\bf 1.00)$,  $\textrm{Het+Serial Correlation}$} \\
		n = 250 & 0.140 & 0.192 & 0.202 & 0.214 & 0.232 & 0.247 & 0.262 & 0.278 & 0.295 & 0.183 \\
		n = 500 & 0.125 & 0.171 & 0.175 & 0.189 & 0.207 & 0.218 & 0.234 & 0.261 & 0.283 & 0.187 \\
		n = 1000 & 0.104 & 0.135 & 0.148 & 0.155 & 0.168 & 0.190 & 0.206 & 0.222 & 0.245 & 0.174 \\
		n = 2000 & 0.104 & 0.120 & 0.128 & 0.141 & 0.150 & 0.155 & 0.170 & 0.195 & 0.210 & 0.174 \\
	\end{tabular}}
	\label{tab:Table4}
\end{table}%


\newpage

\begin{thebibliography}{9}
	

		
	\bibitem[\protect\citeauthoryear{Breitung and Demetrescu}{2015}]{bd2015}
	Breitung, J., and Demetrescu, M. (2015), ``Instrumental variable and variable addition based inference in predictive regressions,'' \textit{Journal of Econometrics}, 187, 358-375.
	
		\bibitem[\protect\citeauthoryear{Elmachkouri2007}{2007}]{elm2007}
	El Machkouri, M., and Ouchti, L. (2007), ``Exact convergence rates in the central limit theorem for a class of martingales,'' \textit{Bernoulli}, 13, 981-999. 
		
	
	\bibitem[\protect\citeauthoryear{Kostakis, Magdalinos and Stamatogiannis}{2015}]{kms2015}
	Kostakis, A., Magdalinos, T., and Stamatogiannis, M. P. (2015), ``Robust econometric inference for stock return predictability,'' \textit{The Review of Financial Studies}, 28, 1506-1553.
	
		\bibitem[\protect\citeauthoryear{Lahiri2009}{2009}]{ls2009}
	Lahiri, S. N., and Sun, S. (2009), ``A Berry-Esseen Theorem for Sample Quantiles under weak dependence,'' \textit{Annals of Applied Probability}, 19, 108-126. 
	
		\bibitem[\protect\citeauthoryear{Magdalinos}{2022}]{tm2022}
	Magdalinos, T. (2022), ``Least squares and IVX limit theory in systems of predictive regressions with GARCH innovations.'', \textit{Econometric Theory}, 38, 875-912.	
	
	\bibitem[\protect\citeauthoryear{Magdalinos and Phillips}{2009}]{mp2009}
	Magdalinos, T., and Phillips, P. C. B. (2009), ``Limit theory for cointegrated systems with moderately integrated and moderately explosive regressors'', \textit{Econometric Theory}, 25, 482-526.
	
		\bibitem[\protect\citeauthoryear{Phillips}{2015}]{pcb2015}
	Phillips, P. C. B. (2015), ``Pitfalls and Possibilities in Predictive Regressions,'' \textit{Journal of Financial Econometrics}, 13, 521-555.
	
	\bibitem[\protect\citeauthoryear{Phillips}{2023}]{pcb2023}
	Phillips, P. C. B. (2023), ``Estimation and inference with near unit roots,'' \textit{Econometric Theory}, 39, 221-263.
	
	\bibitem[\protect\citeauthoryear{Phillips and Magdalinos}{2009}]{pm2009}
	Phillips, P. C. B., and Magdalinos, T. (2009), ``Econometric inference in the vicinity of unity'', Singapore Management University, CoFie Working Paper, 7.	
	
	\bibitem[\protect\citeauthoryear{Yang, Liu, Peng and Cai}{2021}]{ylpc2021}
	Yang, B., X. Liu, L. Peng and Z. Cai (2021), ``Unified Tests for a Dynamic Predictive Regression,'' \textit{Journal of Business and Economic Statistics}, 39, 684-699. 

	
\end{thebibliography}
\end{document}